\documentclass[adp,fleqn,final,finallayout]{w-art}

\usepackage{times}
\usepackage{w-thm}
\usepackage[english]{babel}
\usepackage[nosort]{cite}
\usepackage{graphicx}
\usepackage{color}
\usepackage{amsmath,amssymb,amsfonts}
\usepackage{bm}

\renewcommand{\copyright}{}\copyrightinfo{}{}
\subsectionfont{\normalfont\normalsize\bfseries}
\newcommand{\subsectionmath}[1]{\bm{#1}}

\newcommand{\ket}[1]{|{#1}\rangle}
\newcommand{\bra}[1]{\langle{#1}|}

\newcommand{\floor}[1]{\lfloor{#1}\rfloor}

\newcommand{\IM}{\text{Im}}
\newcommand{\distance}[2]{d_{{#1},{#2}}}
\newcommand{\calA}{{\cal A}}
\newcommand{\calB}{{\cal B}}

\newcommand{\calG}{{\cal G}}
\newcommand{\td}{t_d}
\newcommand{\tone}{{t_1}}

\newcommand{\ttwo}{{t_2}}
\newcommand{\ttwosq}{t_2^2}
\newcommand{\tonetwo}{t_1\text{-}t_2}
\newcommand{\txd}{t_d^{\protect\ast}}
\newcommand{\txone}{{t_1^{\protect\ast}}}
\newcommand{\txonesq}{t_1^{\protect\ast2}}

\newcommand{\txonequ}{t_1^{\protect\ast4}}
\newcommand{\txtwo}{{t_2^{\protect\ast}}}
\newcommand{\txtwosq}{t_2^{\protect\ast2}}

\newcommand{\txtwoqu}{t_2^{\protect\ast4}}
\newcommand{\Woo}{u}                
\newcommand{\Woi}{w}                
\newcommand{\Wio}{\overline{w}}     
\newcommand{\Wii}{v}                
\newcommand{\INF}{^{\infty}}
\newcommand{\bamma}{\bar{\gamma}}
\newcommand{\GA}{\widehat{G}\INF_{\gamma}}
\newcommand{\GB}{\widehat{G}\INF_{\bamma}}
\newcommand{\HNN}{h}
\newcommand{\HNNN}{H}
\newcommand{\Hloc}{H_{\text{loc}}}

\begin{document}

  \DOIsuffix{theDOIsuffix}



  
  \makeatletter\renewcommand{\received@name}{}\makeatother\Receiveddate{~\newline~}



  \title
  {Green functions for nearest- and next-nearest-neighbor hopping on the Bethe lattice}

  \author[M.\ Kollar {\em et al.}]{M.\ Kollar\footnote{Corresponding
      author. E-mail: {\sf Marcus.Kollar@physik.uni-augsburg.de}%
    }\inst{1}}

  \author[]{M.\ Eckstein\inst{1}}

  \author[]{K.\ Byczuk\inst{1,2}}

  \author[]{N.\ Bl\"umer\inst{3}}

  \author[]{P.\ van~Dongen\inst{3}}

  \author[]{M.~H.~Radke de~Cuba\inst{4}}

  \author[]{W.\ Metzner\inst{5}}

  \author[]{D.\ Tanaskovi\'c\inst{6}}

  \author[]{V.\ Dobrosavljevi\'c\inst{6}}

  \author[]{G.\ Kotliar\inst{7}}

  \author[]{D.\ Vollhardt\inst{1}}

  \address[\inst{1}]{Theoretische~Physik~III,
    Elektronische~Korrelationen~und~Magnetismus, Institut~f\"ur~Physik,
    Universit\"at~Augsburg, 86135~Augsburg, Germany}

  \address[\inst{2}]{Institute~of~Theoretical~Physics,
    Warsaw~University, ul.~Ho\.za~69, PL-00-681~Warszawa, Poland}

  \address[\inst{3}]{Institut~f\"ur~Physik, KOMET~337,
    Universit\"at~Mainz, 55099~Mainz, Germany}
  
  \address[\inst{4}]{Werkstra\ss{}e 12, 52076~Aachen, Germany}
  
  \address[\inst{5}]{Max-Planck-Institut~f\"ur~Festk\"orperforschung,
    Heisenbergstr.~1, 70569~Stuttgart, Germany}
  
  \address[\inst{6}]{Department~of~Physics
    and National~High~Magnetic~Field~Laboratory,
    Florida~State~University, 1800~E. Paul~Dirac~Drive, Tallahassee,
    FL~32310-3706, USA}
  
  \address[\inst{7}]{Department~of~Physics~and~Astronomy,
    Rutgers~University, PO~Box~849, Piscataway, NJ~08854-8019,
    USA}
  
  \dedicatory{Dedicated to Bernhard M\"uhlschlegel on the occasion of
    his 80th birthday}

  \begin{abstract}
    We calculate the local Green function for a quantum-mechanical
    particle with hopping between nearest and next-nearest neighbors
    on the Bethe lattice, where the on-site energies may alternate on
    sublattices.  For infinite connectivity the renormalized
    perturbation expansion is carried out by counting all
    non-self-intersecting paths, leading to an implicit equation for
    the local Green function.  By integrating out branches of the
    Bethe lattice the same equation is obtained from a path integral
    approach for the partition function.  This also provides the local
    Green function for finite connectivity.  Finally, a recently
    developed topological approach is extended to derive an operator
    identity which maps the problem onto the case of only
    nearest-neighbor hopping.  We find in particular that hopping
    between next-nearest neighbors leads to an asymmetric spectrum
    with additional van-Hove singularities.
  \end{abstract}
  
  \maketitle

  \renewcommand{\rightmark}{}
  \thispagestyle{headings}

  \section{Introduction}\label{sec:intro}
  
\begin{SCfigure}
    \centering
    \includegraphics[width=0.45\textwidth,angle=0]{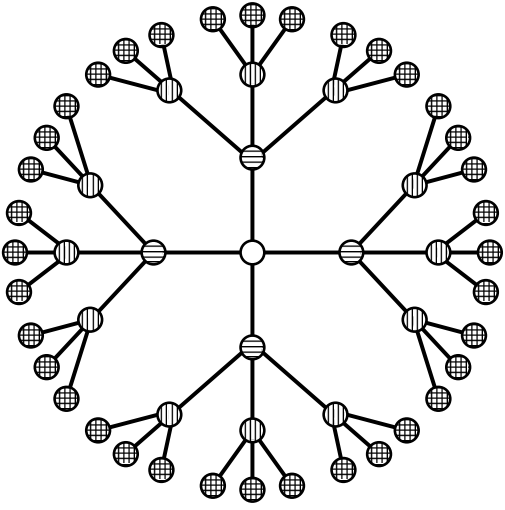}%
    \nopagebreak
    \caption{Part of the Bethe lattice with coordination number $Z=4$. 
      Any two sites are connected by a unique shortest path of bonds.
      Starting from the site marked by the open circle, horizontally
      shaded circles can be reached by one lattice step (NN),
      vertically shaded circles by two lattice steps (NNN), and doubly
      shaded circles by three lattice steps. Note that the lattice is
      infinite and bipartite.}
    \label{fig:bethe}
  \end{SCfigure}

  The Bethe lattice plays an important role in statistical mechanics
  and condensed matter theory.  It is defined as an infinite tree
  graph in which each vertex has $Z$ edges, such that any two vertices
  are connected by only one shortest path, as shown in
  Fig.~\ref{fig:bethe}.  Several physical problems involving
  interactions and/or disorder can be solved exactly for the Bethe
  lattice due to its recursive structure, e.g., Ising
  models\cite{bethe,baxter}, or Anderson localization
  \cite{abouchacra,mirlin,efetov,zirnbauer}.  Furthermore the Bethe
  lattice is useful as a model for the electronic structure of
  amorphous solids\cite{weaire}; see Ref.~\cite{mingo} for a recent
  application.
  
  In this article we study the spectra of tight-binding Hamiltonians
  with hopping between nearest (NN) and next-nearest neighbors (NNN)
  on the Bethe lattice. In this context the standard methods of
  solid-state physics which are based on Bloch's theorem cannot be
  directly applied. This has led to the development of several
  alternative approaches to calculate the tight-binding spectrum for
  NN hopping on the Bethe lattice
  \cite{brinkman-rice,chen,economou,thorpe,mahan,eckstein04}.
  However, some of these methods become very complicated for hopping
  beyond NN due to the proliferation of hopping paths. Below we apply
  and further develop several different calculational schemes to
  determine the Green function in the presence of both NN and NNN
  hopping.
  
  In terms of the quantum-mechanical single-particle operator
  $\ket{i}\bra{j}$, which removes a particle from site $j$ and
  recreates it at site $i$, the general, translationally invariant
  hopping Hamiltonian on the Bethe lattice has the form
  \begin{align}%
    H_{\text{kin}}
    &=
    \sum_{ij}
    t_{ij} \ket{i}\bra{j}
    =
    \sum_{d\geq0}
    \td H_{d}
    \,,\label{eq:H-general}
    &
    H_d
    &=
    \sum_{\distance{i}{j}=d}
    \ket{i}\bra{j}
    \,.
  \end{align}%
  Here hopping processes between two sites $i$ and $j$ are classified
  according to their topological distance $\distance{i}{j}$, i.e., the
  number of nearest-neighbor steps of the shortest path between $i$
  and $j$.  We will also allow for an alternating on-site energy
  $\epsilon_i$,
  \begin{align}
    \Hloc
    &=
    \sum_i
    \epsilon_i \ket{i}\bra{i}
    \,,
    \;\;\;\;\;\;
    \epsilon_i
    =
    \left\{
      \begin{array}{ll}
        \epsilon_{\text{A}}
        &\text{~if~}i\in{\text{A}}
        \\
        \epsilon_{\text{B}}
        &\text{~if~}i\in{\text{B}}        
      \end{array}
    \right.
    \,,\label{eq:H-loc}
  \end{align}
  where $\epsilon_i$ depends only on the sublattice
  $\gamma=\text{A},\text{B}$ of the bipartite Bethe lattice to which
  $i$ belongs. In correlated systems, e.g., for the Hubbard model,
  on-site energies (\ref{eq:H-loc}) may be used to detect
  antiferromagnetic symmetry breaking.
  
  A well-defined limit for infinite coordination number $Z$ results if
  the hopping amplitudes and Hamiltonians are scaled according to
  \cite{vollhardt}
  \begin{align}%
    \td
    &=
    \frac{\txd}{K^{d/2}}
    \,,
    &
    \tilde{H}_{d}
    &=
    \frac{H_{d}}{K^{d/2}}
    \,,
    &
    \td
    H_{d}
    &=
    \txd
    \tilde{H}_{d}    
    \,,\label{eq:scaling}
  \end{align}%
  where $K=Z-1$ is the connectivity.  In the limit $K\to\infty$
  dynamical mean-field theory (DMFT)
  \cite{Georges92,Jarrell92,vollha93,pruschke,georges,PT} becomes
  exact. In particular for the Hubbard model the self-energy is then
  local in space and may be obtained from a single-impurity problem
  with self-consistency condition \cite{georges}.  In recent years
  DMFT for the Hubbard model on the Bethe lattice has greatly helped
  to understand the Mott transition from a paramagnetic metal to a
  paramagnetic insulator at
  half-filling \cite{Georges92,Jarrell92,vollha93,pruschke,%
    georges,PT,bulla99,rozenberg1,Joo,Bulla01,blumer}. For the
  paramagnetic phase to be stable the antiferromagnetic
  low-temperature phase of the Hubbard model needs to be suppressed by
  frustration; only then the Mott transition is observable. This can
  be achieved by including \emph{random} hopping beyond NN
  \cite{georges,rozenberg2,rozenberg3,chitra,zitzler}. In this case the
  density of states (DOS) remains semi-elliptic, implying that the
  Mott transition in the paramagnetic phase is unchanged.  On the
  other hand, for \emph{nonrandom} NNN hopping the DOS is usually
  asymmetric, and this is also the case for the Bethe lattice
  \cite{eckstein04}. Moreover, an asymmetric DOS is known to stabilize
  ferromagnetism away from half-filling
  \cite{wahle98,noack,vollhardt01}.
    
  Here, as a prerequisite for any investigations of frustrated
  interacting or disordered systems, we will study the noninteracting
  Hamiltonian with (nonrandom) NN and NNN hopping
  \begin{align}
    \HNNN
    =
    \Hloc+\tone H_1+\ttwo H_2
    =
    \Hloc+\txone\tilde{H}_1+\txtwo\tilde{H}_2
    \,\label{eq:H-t1t2}
  \end{align}
  and obtain its local Green function $G_{i}(z)$, defined for $\IM\,z\neq0$ by
  \begin{align}
    G_{ij}(z)
    &:=
    \bra{i}\frac{1}{z-H}\ket{j}
    \,,
    &
    G_{i}(z)
    &:=
    G_{ii}(z)
    \,,
    &
    G\INF_i(z)
    &:=
    \lim_{K\to\infty}G_i(z)
    \,,\label{eq:G-definition}
  \end{align}
  paying special attention to the limit $K\to\infty$.  Note that due
  to translational symmetry of the infinite Bethe lattice $G_{i}(z)=:G_{\gamma}(z)$ depends only on
  the sublattice $\gamma$ of $i$.  We recall that for only NN hopping
  the local Green function for sublattice $\gamma$ is given by
  \cite{economou}
  \begin{align}
    g_{\gamma}(z)
    &:=
    \bra{i}\frac{1}{z-h}\ket{i}    
    =
    \frac{
      2K(z-\epsilon_{\bamma})
    }{
      (K-1)x+(K+1)\sqrt{x-4\txonesq}\sqrt{x}
    }
    \,,
    &
    \HNN
    &=
    \Hloc+\txone\tilde{H}_1
    \,,\label{eq:nn-only}
  \end{align}
  where $x=(z-\epsilon_{\text{A}})(z-\epsilon_{\text{B}})$ and the
  square roots are given by their principal branches.
  
  The derivation of the local Green function for $\tonetwo$ hopping
  will proceed as follows.  In Sec.~\ref{sec:rpe} we use the
  renormalized perturbation expansion (RPE) \cite{economou} to obtain
  a closed set of equations for $G\INF_{\gamma}(z)$, which are also
  related to the DMFT selfconsistency equations. The RPE method is
  well-suited for the Bethe lattice due to its recursive nature,
  although the classification of paths for $\ttwo\neq0$ is rather
  involved. We show how to use the RPE result to establish the
  asymmetry of the DOS.
  On the other hand, in Sec.~\ref{sec:pathintegrals} we use the
  many-body path integral approach \cite{negeleorland} to derive an
  exact effective action by a recursive method. This also yields
  closed equations for the local Green functions $G_{\gamma}(z)$ for
  any coordination number $Z$.  Furthermore a surprising algebraic
  relation between Green functions for finite and infinite $Z$ is
  uncovered.
  Finally, in Sec.~\ref{sec:operatormethod} a recently developed
  topological method \cite{eckstein04} is extended to include $\Hloc$.
  We derive an operator identity, valid for any $Z$, that allows one
  to express $G_{\gamma}(z)$ in terms of the known solutions
  (\ref{eq:nn-only}) for only NN hopping.
  Our results for the local Green function are discussed in
  Sec.~\ref{sec:green}. A conclusion in Sec.~\ref{sec:conclusion}
  closes the presentation.

  \section{Renormalized perturbation expansion}\label{sec:rpe}
  
  \subsection{Dressed expansion for the local Green function}
  
  In this section we obtain an equation for the local Green function
  (\ref{eq:G-definition}) using the renormalized perturbation
  expansion (RPE) \cite{economou,vlaming92}. In this approach the
  Green function $G_{ij}(z)$ for $\HNNN$ is obtained in terms of the
  Green function
  $G_{ij}^{\text{loc}}(z)=\delta_{ij}G_i^{\text{loc}}(z)$ for $\Hloc$
  as
  \begin{align}
    G_{ij}
    &=
    \delta_{ij}G_i^{\text{loc}}
    +
    G_i^{\text{loc}}t_{ij}G_j^{\text{loc}}
    +
    \sum_kG_i^{\text{loc}}t_{ik}G_k^{\text{loc}}t_{kj}G_j^{\text{loc}}
    +
    \cdots
    \,,
    &
    G_i^{\text{loc}}
    &=
    \frac{1}{z-\epsilon_i}
    \,,
  \end{align}
  where we omit the argument $z$ for the moment.  This expansion
  contains terms in which site indices are repeated. In a graphical
  representation this corresponds to paths in which some lattice sites
  are ``decorated'' with closed paths \cite{economou}. One can omit
  these decorations at the first site $i$ by replacing
  $G_i^{\text{loc}}$ by the full local Green function $G_i$.  At the
  next site $k$, however, the decorations at site $i$ must not be
  repeated, leading to the replacement of $G_k^{\text{loc}}$ by the
  local Green function with site $i$ removed, i.e., by
  $G_{k}^{[i]}=G_{k}|_{\epsilon_i=\infty}$. Repeating this process one
  obtains
  \begin{align}
    G_i
    &=
    G_i^{\text{loc}}
    +
    {\sum_k}'
    G_i\,t_{ik}\,G_{k}^{[i]}\,t_{ki}\,G_i^{\text{loc}}
    +
    {\sum_{km}}'
    G_i\,t_{ik}\,G_{k}^{[i]}\,t_{km}\,G_{m}^{[i,k]}\,t_{mi}\,G_i^{\text{loc}}
    +
    \cdots
    \,,
  \end{align}
  where the primed sums are now only over non-self-inter\-secting
  paths.  The RPE is particularly useful in the limit $Z\to\infty$
  which allows the replacement $G_i^{[\cdots]}\to G\INF_i$.  This
  yields an equation involving only local Green functions,
  \begin{align}
    G\INF_i(z)^{-1}
    &=
    z-\epsilon_i
    -
    \Bigg[
    {\sum_k}
    t_{ik}\,G\INF_{k}(z)\,t_{ki}
    +
    {\sum_{km}}'
    t_{ik}\,G\INF_{k}(z)\,t_{km}\,G\INF_{m}(z)\,t_{mi}
    + 
    \cdots
    \Bigg]
    \,.\label{eq:rpe-infty}
  \end{align}
  For the case of two sublattices with on-site energies
  $\epsilon_{\text{A}}$, $\epsilon_{\text{B}}$ we thus obtain two
  coupled equations ($\gamma=\text{A,B}$; $\bar{\text{A}}=\text{B}$, $\bar{\text{B}}=\text{A}$),
  \begin{align}
    G\INF_{\gamma}(z)^{-1}
    &=
    z-\epsilon_{\gamma}-F(G\INF_{\gamma}(z),G\INF_{\bamma}(z))
    \,.\label{eq:rpe-F-expression}
  \end{align}
  This is a closed system of implicit equations for the local Green functions
  $G\INF_{\gamma}(z)$.
  
  Note that the self-consistency equations of DMFT are essentially
  contained in Eq.~(\ref{eq:rpe-F-expression}). In DMFT the
  self-energy is local \cite{georges},
  $\Sigma_{ij}(z)=\Sigma_{\gamma}(z)\delta_{ij}$ for $i\in\gamma$, and
  the local \emph{interacting} Green function
  $G^{\text{int}}_{\gamma}(z)$ is given by the Dyson equation
  $G^{\text{int}}_{\gamma}(z)^{-1} =
  G\INF_{\gamma}(z-\Sigma_{\gamma}(z))^{-1} =
  \calG_{\gamma}(z)^{-1}-\Sigma_{\gamma}(z)$, where
  $\calG_{\gamma}(z)$ is the Weiss field of an auxiliary single-site
  problem for sublattice $\gamma$. From
  Eq.~(\ref{eq:rpe-F-expression}) one thus obtains the
  self-consistency equation
  \begin{align}
    \calG_{\gamma}(z)^{-1}
    &=
    z-\epsilon_{\gamma}-F(G^{\text{int}}_{\gamma}(z),G^{\text{int}}_{\bamma}(z))
    \,.\label{eq:DMFT-sc}
  \end{align}
  Here spin indices were suppressed for simplicity. Detailed
  discussions of DMFT self-consistency equations can be found in
  Refs.~\cite{georges,eckstein04}.

  \subsection{RPE for $\subsectionmath{\tonetwo}$ hopping on the Bethe lattice}
  
  For the remainder of this section we consider the Bethe lattice in
  the limit $Z\to\infty$.  As a standard example, we first consider
  only NN hopping (\ref{eq:nn-only}).  Since no closed loops are
  possible, the only allowed non-self-intersecting paths are visits to
  one of the $Z$ NN sites which return immediately. This yields the
  well-known equation \cite{economou}
  \begin{align}
    g\INF_{\gamma}(z)^{-1}
    &=
    z-\epsilon_{\gamma}-\txonesq\,g\INF_{\bamma}(z)
    \,,
    &
    g\INF_{\gamma}(z)
    &=
    \frac{
      2(z-\epsilon_{\bamma})
    }{
      x+\sqrt{x-4\txonesq}\sqrt{x}
    }
    \,,
    &
    (\ttwo&=0)
  \end{align}
  where the solution is a special case of Eq.~(\ref{eq:nn-only}).
  
\begin{figure}[tbp]
    (a)\includegraphics[scale=0.45,angle=0]{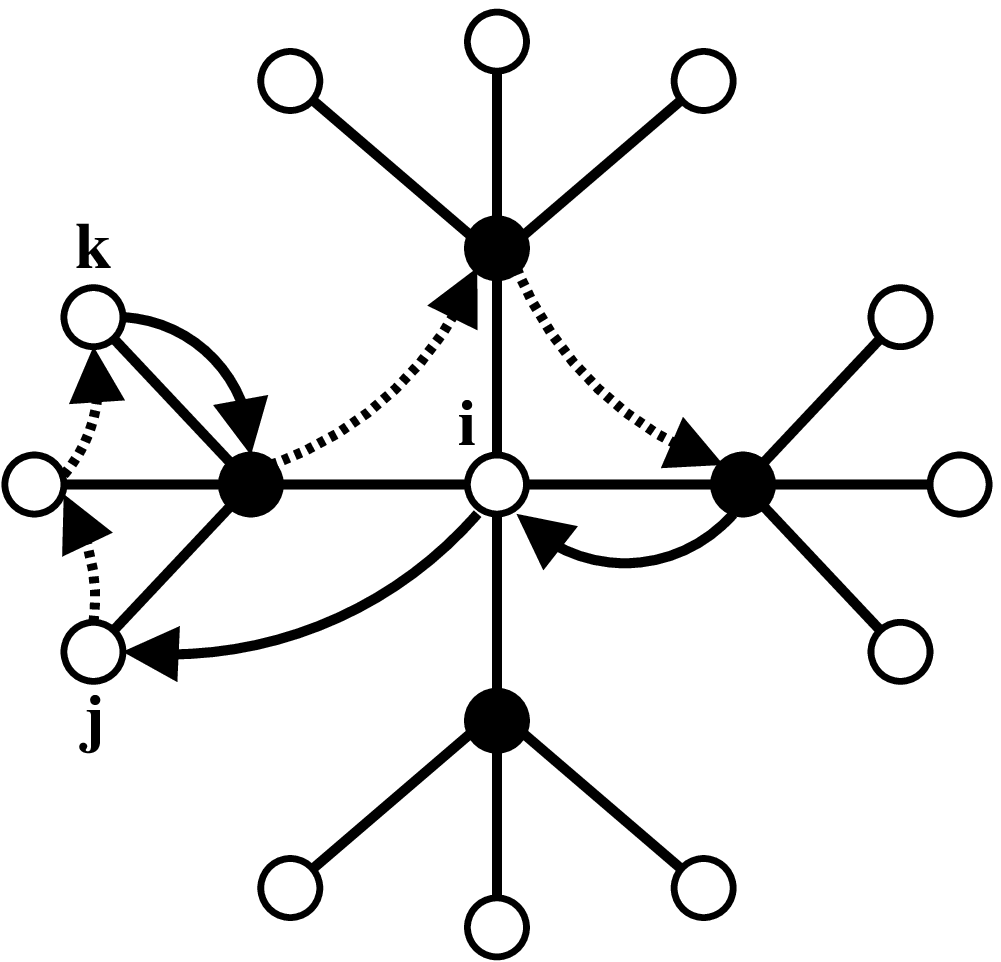}%
    \hfill%
    (b)\includegraphics[scale=0.45,angle=90]{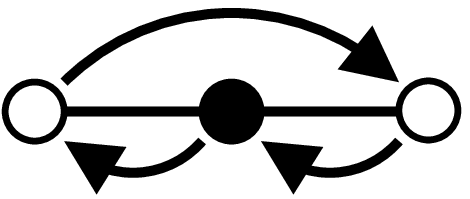}%
    \hfill%
    (c)\includegraphics[scale=0.45,angle=0]{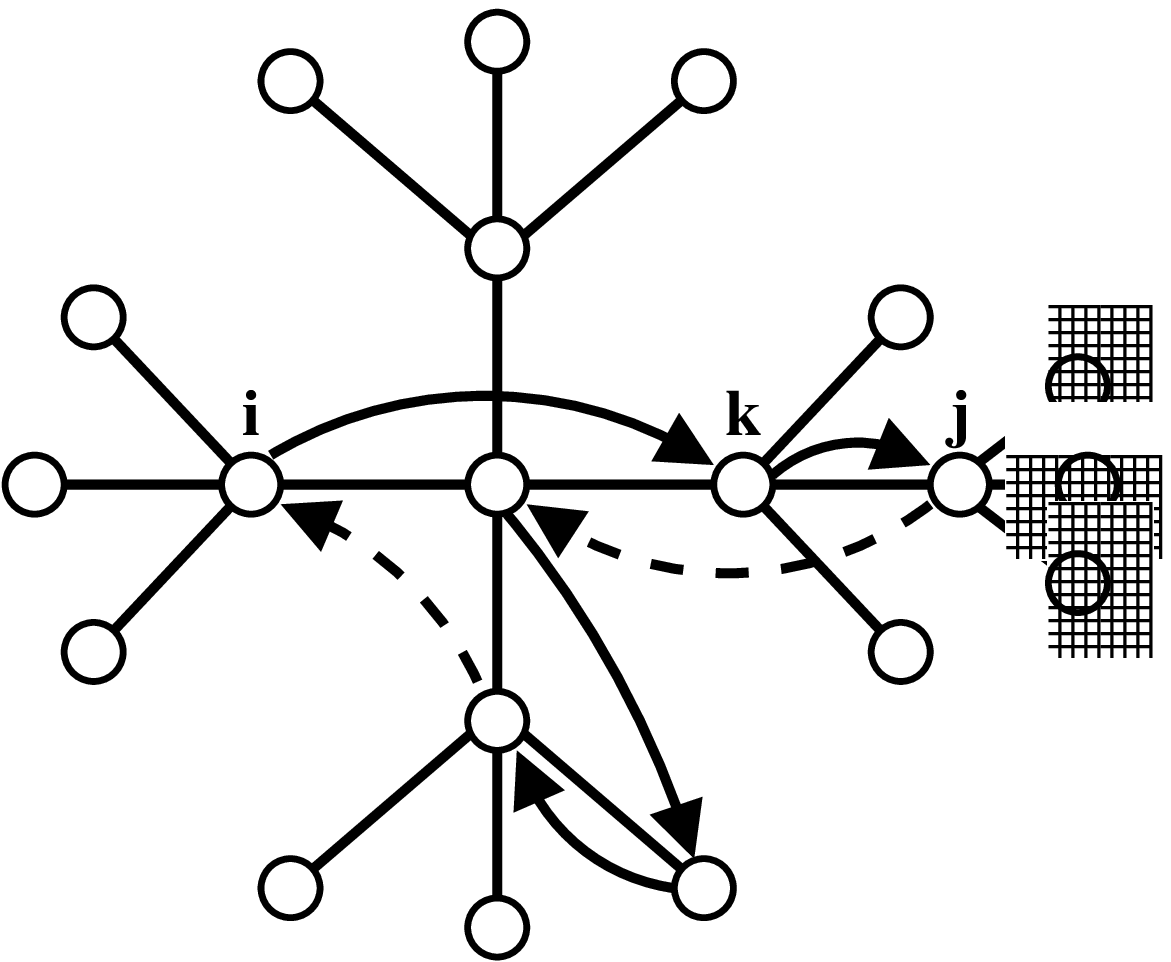}%
    \caption{(a)~Example of intra-shell detours in RPE.
      A non-self-intersecting path, which starts and ends at site $i$,
      may contain NNN steps (dotted) to sites with the same distance
      from $i$, e.g., from $j$ to $k$.  Open and solid circles
      indicate $\gamma$ and $\bamma$ sublattices. (b)~Simplified
      diagram for the proper path corresponding to~(a).
      (c)~Rule~(\ref{eq:rpe-rule1}) for proper paths: After visiting
      $k$ and $j$, the shaded sites are blocked and one must return
      towards $i$.  From a NN site of $i$ it may then either return
      immediately to $i$ or leave again with NNN steps, eventually
      taking a NN step and being forced back to $i$.  Dashed arrows
      indicate forced moves.}
    \label{fig:rpe-rules}    
  \end{figure}

  We now proceed to the case of $\tonetwo$ hopping, for which the
  evaluation of the square bracket in Eq.~(\ref{eq:rpe-infty}) is more
  involved, since it requires the enumeration \cite{radke,blumer} of
  several classes of closed non-self-intersecting paths which begin
  and end at site $i$.  First we note that at each site $j$ ($\neq i$)
  on the path it is possible to make a detour within the same shell,
  i.e., to one of the yet unvisited NNN sites $k$ of $j$ with same
  distance $\distance{i}{j}=\distance{i}{k}$ from $i$, as illustrated
  in Fig.~\ref{fig:rpe-rules}a.  In the limit $Z\to\infty$ we can take
  these detours into account by replacing $G\INF_{\gamma}$ by
  \begin{align}
    \GA
    =
    G\INF_{\gamma}
    +
    G\INF_{\gamma}\,\txtwo\,G\INF_{\gamma}
    +
    G\INF_{\gamma}\,\txtwo\,G\INF_{\gamma}\,\txtwo\,G\INF_{\gamma}
    +
    \cdots
    =
    \frac{G\INF_{\gamma}}{1-\txtwo\,G\INF_{\gamma}}
    \label{eq:rpe-Ghat-definition}
  \end{align}
  and only considering non-self-intersecting paths that \emph{change
    shells in every step}. Such shell-changing, non-self-intersecting
  paths are referred to as \emph{proper paths} from now on. They may
  be drawn using simplified diagrams indicating the visited shells
  only, see Fig.~\ref{fig:rpe-rules}b. Closed proper paths, starting
  and ending at site $i$, are then governed by the following rule:
  \begin{align}%
    \parbox[t]{0.75\textwidth}{Whenever two neighboring sites $j$ and
      $k$ have been visited, with $j$ further away from $i$ than $k$,
      all sites on $j$'s branch further away from $i$ than $j$ cannot
      be visited.}
    \label{eq:rpe-rule1}
  \end{align}%
  This rule, which is illustrated in Fig.~\ref{fig:rpe-rules}c, is due
  to the tree-like structure of the Bethe lattice and the requirements
  that the paths are non-self-intersecting, involve only NN or NNN
  steps, and change shells in every step. It is not affected by
  intra-shell detours contained in $\GA$.

\begin{figure}[tbp]
    \begin{tabular*}{\textwidth}{@{\extracolsep{\fill}}c@{}c@{}c@{}c@{}c@{}}
      \includegraphics[scale=0.45,angle=90]{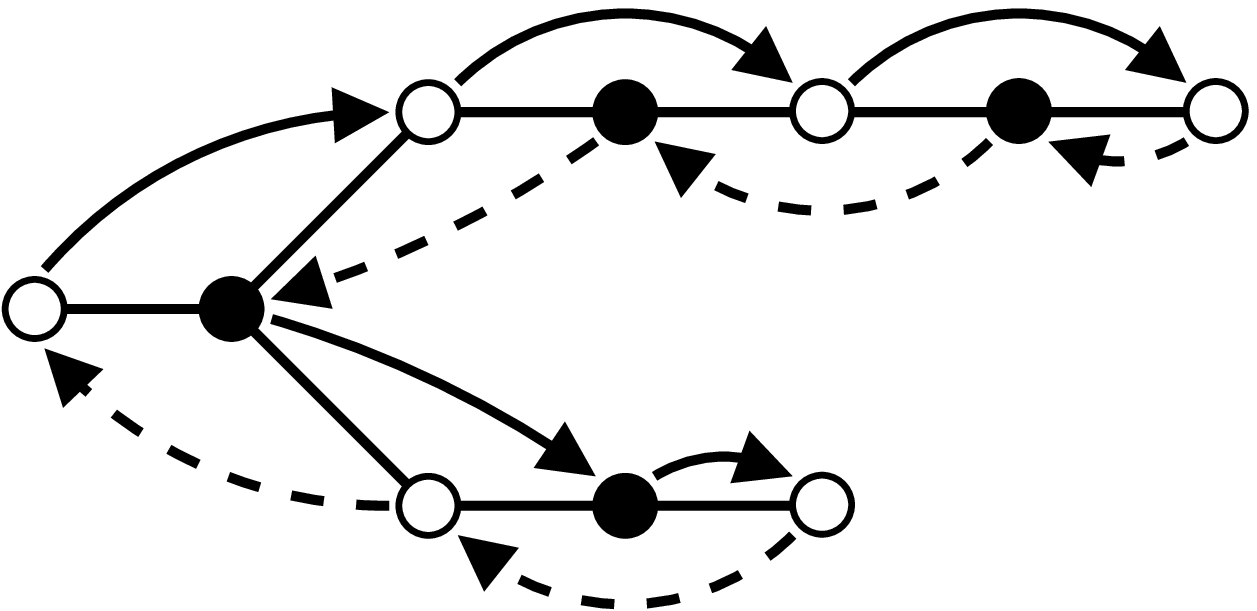} &
      \includegraphics[scale=0.45,angle=90]{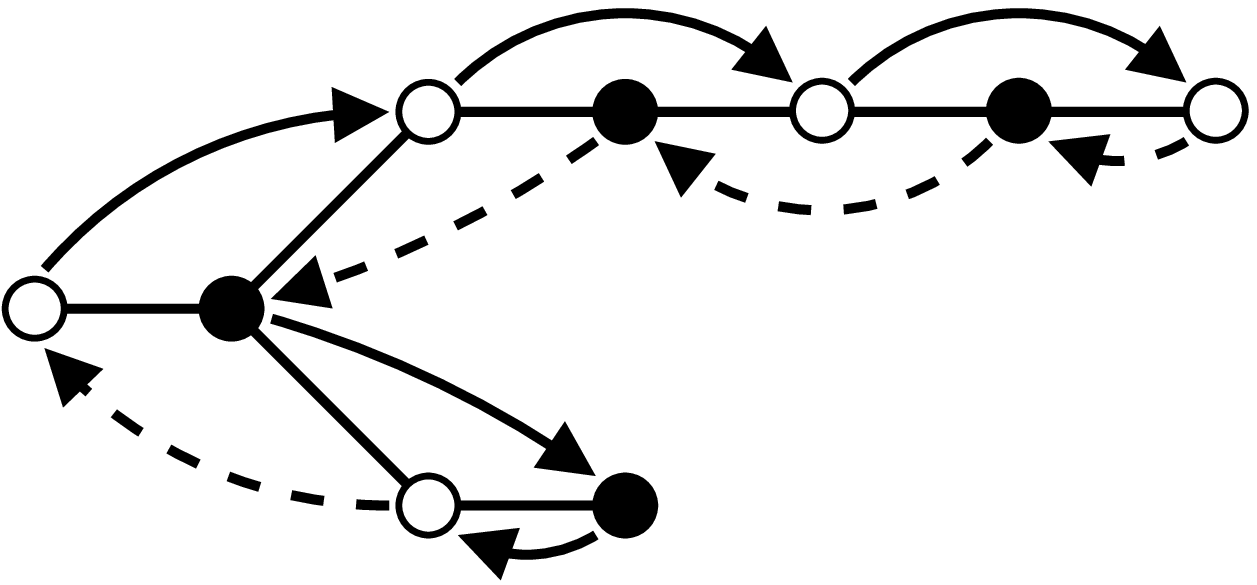} &
      \includegraphics[scale=0.45,angle=90]{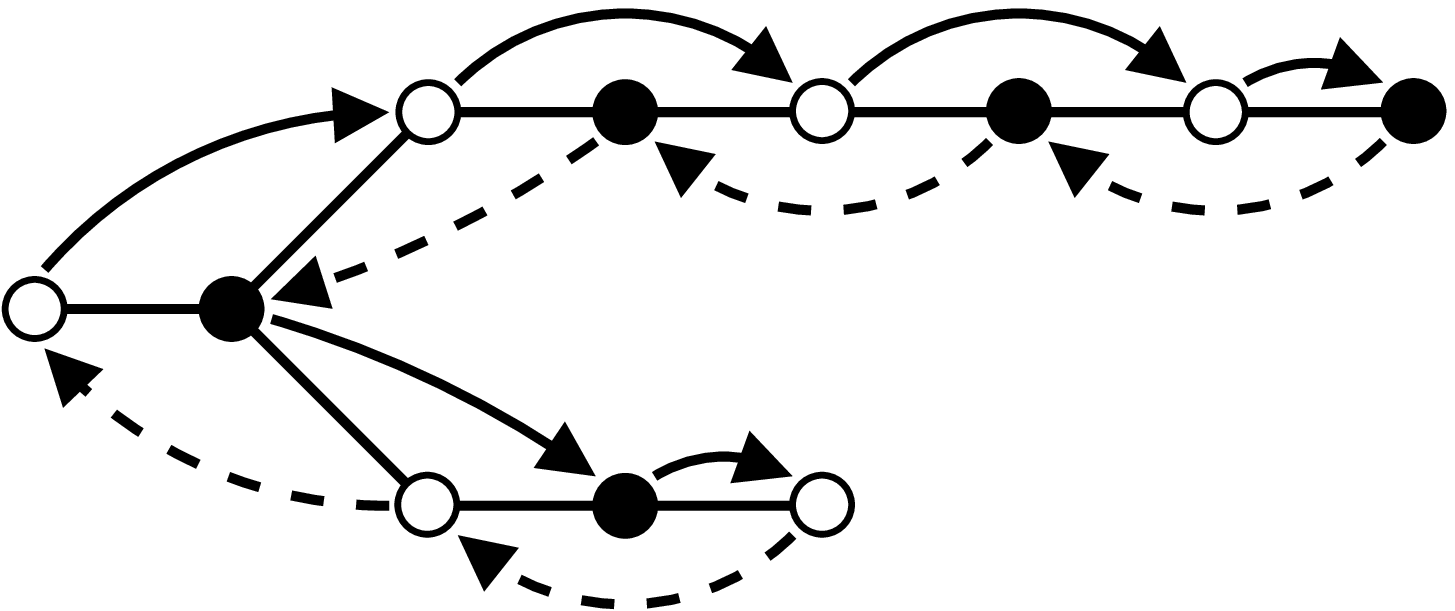} &
      \includegraphics[scale=0.45,angle=90]{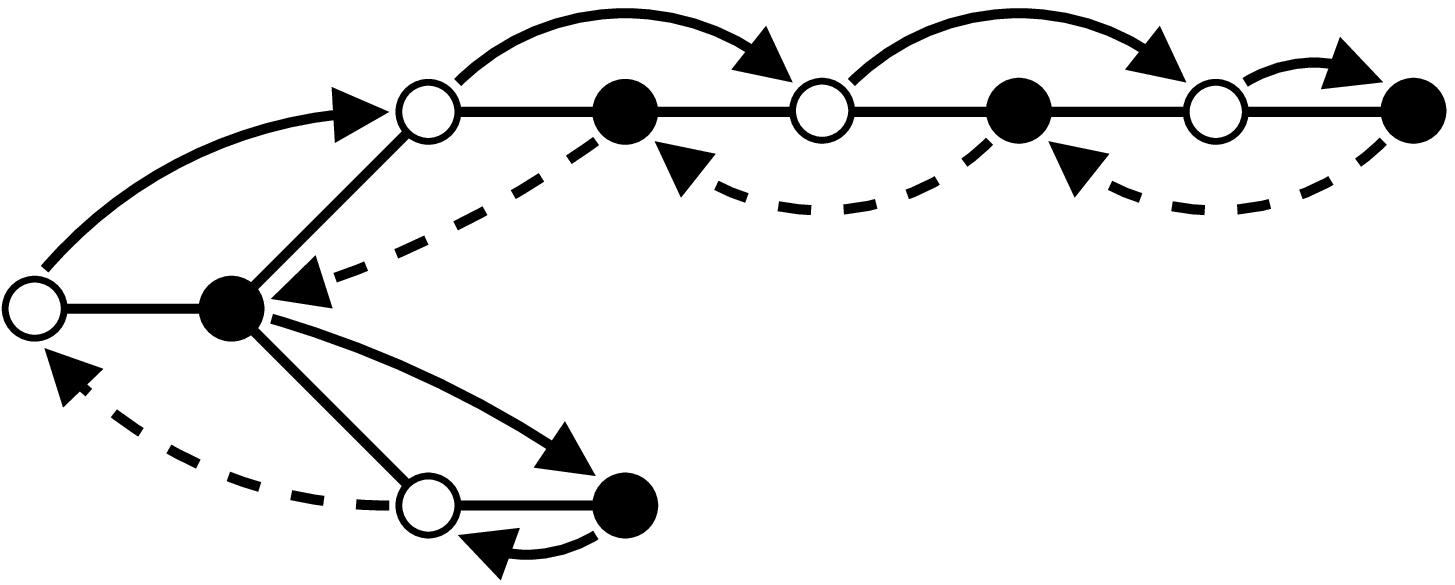} &
      \includegraphics[scale=0.45,angle=90]{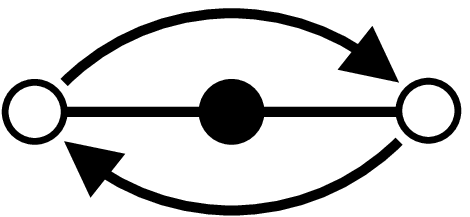}
      \\[2ex]
      (a1) &
      (a2) &
      (b1) &
      (b2) &
      (c)
    \end{tabular*}
    \caption{Simplified diagrams for proper paths that contribute
      to Eq.~(\ref{eq:rpe-F-result}). Only paths which
      change the shell in each step are considered; the effect of 
      intra-shell detours is taken into account by 
      working with dressed Green functions $\GA$, $\GB$. 
      Open and solid circles indicate
      $\gamma$ and $\bamma$ sublattices, respectively. Dashed arrows indicate moves
      that are forced due to the rules (\ref{eq:rpe-rule1}). See
      Table~\ref{tab:rpe} for details.}
    \label{fig:rpe} 
  \end{figure}

  Therefore a sequence of outgoing NNN steps, starting at a site in
  shell $0$ or $1$ and terminated by a NN step \emph{must} be
  followed by a NNN step going inward until shell $1$ or $0$,
  respectively, is reached.  We are thus led to the following
  classification of proper paths by their outermost shell
  $s_{\text{max}}$, as shown in Fig.~\ref{fig:rpe} and listed in
  Table~\ref{tab:rpe}:

  \begin{itemize}
  \item[(a)] An even shell $s_{\text{max}}$ is reached by taking NNN
    steps.  After taking one NN step inward the path must return
    towards $i$ until reaching shell $1$ due to rule
    (\ref{eq:rpe-rule1}). Then the path may turn outward again, finally using
    a NN step (a1) inward or (a2) outward to reach $i$'s sublattice
    and return. The reverse path is also possible, but in case (a1)
    may be identical.
  \item[(b)] An odd shell $s_{\text{max}}$ is reached by taking NNN
    steps and one NN step outward. After returning to shell $1$ due to
    rule (\ref{eq:rpe-rule1}), one may turn outward again in (b1) and
    (b2) similar to (a1) and (a2). The reverse path is also possible,
    but in case (b2) may be identical.
  \item[(c)] Whereas (a) and (b) involve at least one NN step, there
    is also one proper path involving no NN step; it visits a NNN site
    and returns immediately.
  \end{itemize}
  Note that paths that start with a NN step are included in the
  inverted paths mentioned under (a) and (b).  The contributions of
  all proper paths are also collected in Table~\ref{tab:rpe}, and
  their sum is
  \begin{align}
    F(G\INF_{\gamma},G\INF_{\bamma})
    &=
    \sum_{m=1}^{\infty} \sum_{k=0}^{m-1} \left[ G^{\text{(a1)}}_{\gamma,m,k} + G^{\text{(a2)}}_{\gamma,m,k} \right]
    +
    \sum_{m=0}^{\infty} \sum_{k=0}^{m}   \left[ G^{\text{(b1)}}_{\gamma,m,k} + G^{\text{(b2)}}_{\gamma,m,k} \right]
    +
    G^{\text{(c)}}_{\gamma}
    \nonumber\\
    &=
    \frac{\txonesq\,G\INF_{\bamma}\,(1-\txtwo\,G\INF_{\bamma})}{(1-\txtwo\,G\INF_{\gamma}-\txtwo\,G\INF_{\bamma})^2}
    +
    \frac{\txtwosq\,G\INF_{\gamma}}{1-\txtwo\,G\INF_{\gamma}}
    \,.\label{eq:rpe-F-result}
  \end{align}
  Eqs.~(\ref{eq:rpe-F-expression}) and (\ref{eq:rpe-F-result}) are a
  set of coupled quartic equations for $G\INF_{\gamma}$ and
  $G\INF_{\bamma}$.  Eq.~(\ref{eq:rpe-F-result}), taken together with
  (\ref{eq:DMFT-sc}), fully determines the DMFT self-consistency
  equation for $\tonetwo$ hopping on the Bethe lattice.  This result
  differs \cite{radke,eckstein04} from the self-consistency equations
  employed in Refs.~\cite{georges,rozenberg2,rozenberg3,chitra,zitzler},
  which apply only to random hopping.

  \begin{table}[p]
    \begin{tabular*}{\textwidth}{@{\extracolsep{\fill}}l@{}l@{}l@{}l@{}l@{}}
        & path & steps                                      & factor                     & reached shell
        \\
        \hline
        \hline
        \\
        (a)   & $s_{\text{max}}=2m$
               & 1. $m$ NNN steps outward                   & $(\txtwo)^m(\GA)^m$         & $2m$  
        \\
              & ($m\geq1$)
               & 2. $1$ NN step inward                      & $(\txone)(\GB)$             & $2m-1$  
        \\
              && 3. $(m-1)$ NNN steps inward$^{\ast}$       & $(\txtwo)^{m-1}(\GB)^{m-1}$ & $1$  
        \\
              && 4. $k$ NNN steps outward ($k\leq m-1$)     & $(\txtwo)^k(\GB)^k$         & $2k+1$  
        \\[1ex] \cline{2-5} \\[-1ex]
        & (a1) & 5. $1$ NN step outward                     & $(\txone)(\GA)$             & $2k+2$  
        \\
              && 6. $(k+1)$ NNN steps inward$^{\ast}$       & $(\txtwo)^{k+1}(\GA)^{k}$   & $0$  
        \\
              && reverse path (if different)                & $(2-\delta_{k,m-1})$        &
        \\[1ex]
              && \multicolumn{3}{l}{\fbox{$G^{\text{(a1)}}_{\gamma,m,k}=(2-\delta_{k,m-1})(\txone)^2(\txtwo)^{2m+2k}(\GA)^{m+k+1}(\GB)^{m+k}$}}
        \\[2ex] \cline{2-5} \\[-1ex]
        & (a2) & 5. $1$ NN step inward                      & $(\txone)(\GA)$             & $2k$  
        \\
              && 6. $k$ NNN steps inward$^{\ast}$           & $(\txtwo)^{k}(\GA)^{k-1}$   & $0$  
        \\
              && reverse path                               & $2$                         &
        \\[1ex]
              && \multicolumn{3}{l}{\fbox{$G^{\text{(a2)}}_{\gamma,m,k}=2(\txone)^2(\txtwo)^{2m+2k-1}(\GA)^{m+k}(\GB)^{m+k}$}}
        \\[2ex] \hline \\[-1ex]
        (b)   & $s_{\text{max}}=2m+1$
               & 1. $m$ NNN steps outward                   & $(\txtwo)^m(\GA)^m$         & $2m$  
        \\
              & ($m\geq0$)
               & 2. $1$ NN step outward                     & $(\txone)(\GB)$             & $2m+1$  
        \\
              && 3. $m$ NNN steps inward$^{\ast}$           & $(\txtwo)^{m}(\GB)^{m}$     & $1$  
        \\
              && 4. $k$ NNN steps outward ($k\leq m$)       & $(\txtwo)^k(\GB)^k$         & $2k+1$  
        \\[1ex] \cline{2-5} \\[-1ex]
        & (b1) & 5. $1$ NN step outward (if $k\neq m)$      & $(\txone)(\GA)$             & $2k+2$  
        \\
              && 6. $(k+1)$ NNN steps inward$^{\ast}$       & $(\txtwo)^{k+1}(\GA)^{k}$   & $0$  
        \\
              && reverse path                               & $2$                         &
        \\[1ex]
              && \multicolumn{3}{l}{\fbox{$G^{\text{(b1)}}_{\gamma,m,k}=2(1-\delta_{k,m})(\txone)^2(\txtwo)^{2m+2k+1}(\GA)^{m+k+1}(\GB)^{m+k+1}$}}
        \\[2ex] \cline{2-5} \\[-1ex]
        & (b2) & 5. $1$ NN step inward                      & $(\txone)(\GA)$             & $2k$  
        \\
              && 6. $k$ NNN steps inward$^{\ast}$           & $(\txtwo)^{k}(\GA)^{k-1}$   & $0$  
        \\
              && reverse path (if different)                & $(2-\delta_{k,m})$          &
        \\[1ex]
              && \multicolumn{3}{l}{\fbox{$G^{\text{(b2)}}_{\gamma,m,k}=(2-\delta_{k,m})(\txone)^2(\txtwo)^{2m+2k}(\GA)^{m+k}(\GB)^{m+k+1}$}}
        \\ [2ex] \hline \\[-1ex]
        (c)   & $s_{\text{max}}=2$
               & 1. $1$ NNN step outward                    & $(\txtwo)(\GA)$             & $2$  
        \\
              & (no NN steps)
               & 2. $1$ NN step inward                      & $(\txtwo)$                  & $0$  
        \\[1ex]
              && \multicolumn{3}{l}{\fbox{$G^{\text{(c)}}_{\gamma}=(\txtwo)^{2}(\GA)$}}
        \\[2ex] \hline
      \end{tabular*}
    \caption{Proper paths contributing to Eq.~(\ref{eq:rpe-F-result})
      for a Bethe lattice with $Z\to\infty$. Paths (a1) 
      and (a2) reach an even outermost shell, while paths (b1) and (b2) 
      reach an odd outermost shell; both contain at least one
      NN step. On the other hand, without NN steps only path (c) 
      is possible. Steps forced by rule (\ref{eq:rpe-rule1})
      are marked by an asterisk~($^{\ast}$). The reversion of each
      non-symmetric path yields a factor of $2$.} 
    \label{tab:rpe}
  \end{table}  

  The result (\ref{eq:rpe-F-result}) can immediately be used to
  determine the moments $M_n$ of the density of states. For the case
  $\epsilon_{A}=\epsilon_{B}=0$ we have the large-$z$ expansion
  \begin{align}
    G\INF_{\gamma}(z)
    &=
    \int\limits_{-\infty}^{\infty}
    \frac{\rho\INF(\epsilon)}{z-\epsilon}
    \,d\epsilon
    =:
    \sum_{n=0}^{\infty}
    \frac{M_n}{z^{n+1}}
    \,,
    &
    \rho\INF(\epsilon)
    &:=
    -\frac{1}{\pi}\IM\,G\INF_{\gamma}(\epsilon+i0)
    \,.\label{eq:rpe-moments}
  \end{align}
  Multiplying Eq.~(\ref{eq:rpe-F-expression}) by $G\INF_{\gamma}$,
  inserting Eq.~(\ref{eq:rpe-moments}), and comparing coefficients of
  powers of $z$ we find
  \begin{align}
    M_1
    &=
    0
    \,,&
    M_2
    &=
    \txonesq+\txtwosq
    \,,&
    M_3
    &=
    (3\txonesq+\txtwosq)\,\txtwo
    \,,&
    M_4
    &=
    2\txonequ+12\txonesq\txtwosq+3\txtwoqu
    \,,
  \end{align}
  revealing at once that the DOS $\rho\INF(\epsilon)$ is asymmetric
  for $\txtwo\neq0$ \cite{blumer}, in contrast to the case of random
  hopping \cite{georges,rozenberg2,rozenberg3,chitra,zitzler}.
  
  Clearly the RPE is well-suited for the Bethe lattice because rule
  (\ref{eq:rpe-rule1}) represents a strict constraint on proper paths.
  However, already for the case of $\tonetwo$ hopping the RPE requires
  some care.  Furthermore, the enumeration of paths for hopping
  \emph{beyond NNN}, or for $Z<\infty$, is likely to be very tedious.
  Another drawback is that the RPE only yields an implicit system of
  equations for the local Green functions.  Below we derive explicit
  expressions for them via a different route.

  \section{Path-integral approach}\label{sec:pathintegrals}

  \subsection{Green function as path integral}
  
  In this section we use the standard many-body path-integral approach
  to the local Green function.  For non-interacting spinless fermions
  with Hamiltonian $\Hloc+H_{\text{kin}}$ the action is
  \cite{negeleorland}
  \begin{align}
    S
    &=
    \int_0^{1/T}
    \sum_{ij}
    \overline{c}_i(\tau)
    \left[
      (\partial_{\tau}-\mu+\epsilon_i)\delta_{ij}
      +
      t_{ij}
    \right]
    c_j(\tau)
    \;
    d\tau
    \,,
  \end{align}
  where $\overline{c}_i(\tau)$, $c_i(\tau)$ are Grassmann variables.
  Fourier transforming to fermionic Matsubara frequencies, $\omega_n$
  $=$ $(2n+1)\pi T$, we obtain $S$ = $\sum_n \sum_{ij}
  \left[t_{ij}+(\epsilon_i-\mathrm{i}\omega_n-\mu)\delta_{ij}\right]
  \overline{c}_{in}c_{jn}$, i.e., the functional integral factorizes
  with respect to $n$.  We thus consider a fixed Matsubara frequency,
  set $z$ $=$ $\mathrm{i}\omega_n+\mu$, and omit the index $n$.  The
  local Green function is then given by
  \begin{align}
    G_{i}(z)
    &=
    \int\prod_{j}{\cal D}[\overline{c}_{j},c_{j}]
    \;
    e^{-\tilde{S}}
    \;
    \overline{c}_{i}c_{i}
    ~~\Big/~~
    \int\prod_{j}{\cal D}[\overline{c}_{j},c_{j}]
    \;
    e^{-\tilde{S}}
    \,,\label{eq:G-pathint1}
  \end{align}
  with $\tilde{S}$ $=$ $\sum_{i}
  S_{\text{loc}}(i)+\frac{1}{2}\sum_{ij}S_{\text{hop}}(i,j)$,
  $S_{\text{loc}}(i)$ $=$ $-(z-\epsilon_i)\overline{c}_{i}c_{i}$,
  $S_{\text{hop}}(i,j)$ $=$
  $t_{ij}(\overline{c}_{i}c_{j}+\overline{c}_{j}c_{i})$.  Note that
  for this non-interacting system the many-body Matsubara Green
  function is independent of the temperature $T$ and coincides with
  the Green function defined in Eq.~(\ref{eq:G-definition}).

  \subsection{Decomposition of the action for $\subsectionmath{\tonetwo}$ hopping on the Bethe lattice}
  
\begin{figure}[b]
    (a)\includegraphics[width=0.3\textwidth,angle=0]{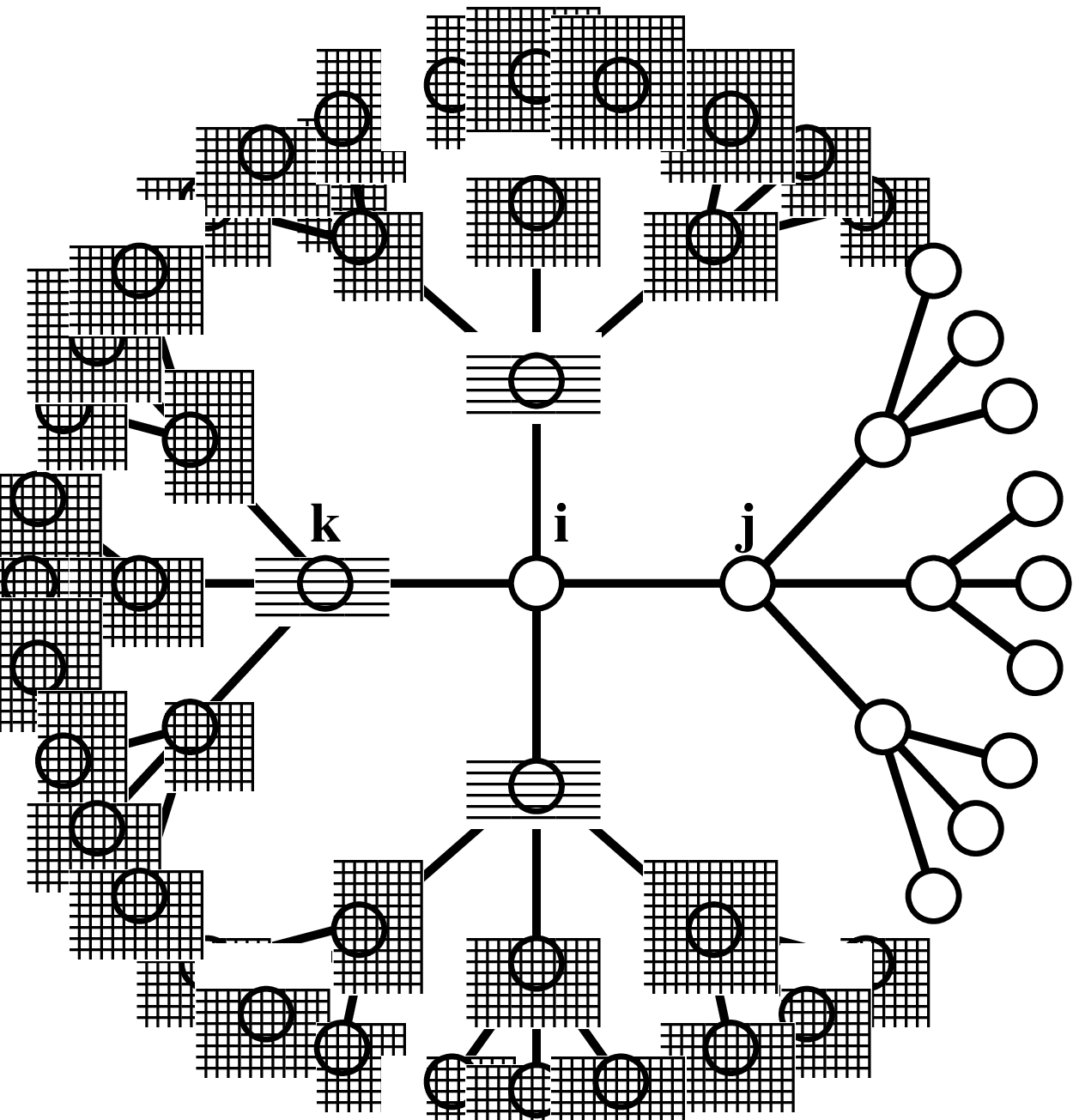}\hfill
    (b)\includegraphics[width=0.3\textwidth,angle=0]{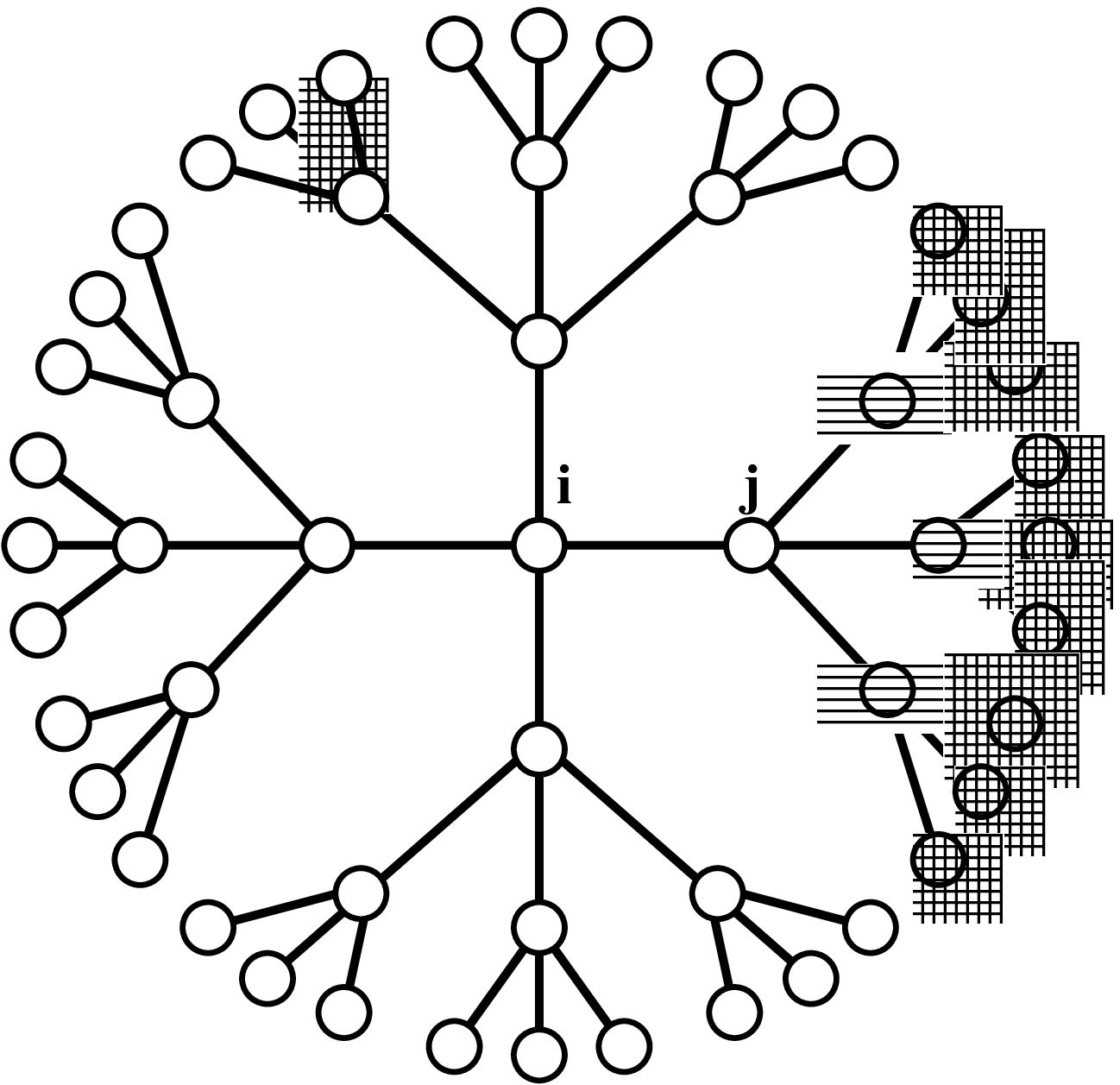}\hfill
    (c)\includegraphics[width=0.3\textwidth,angle=0]{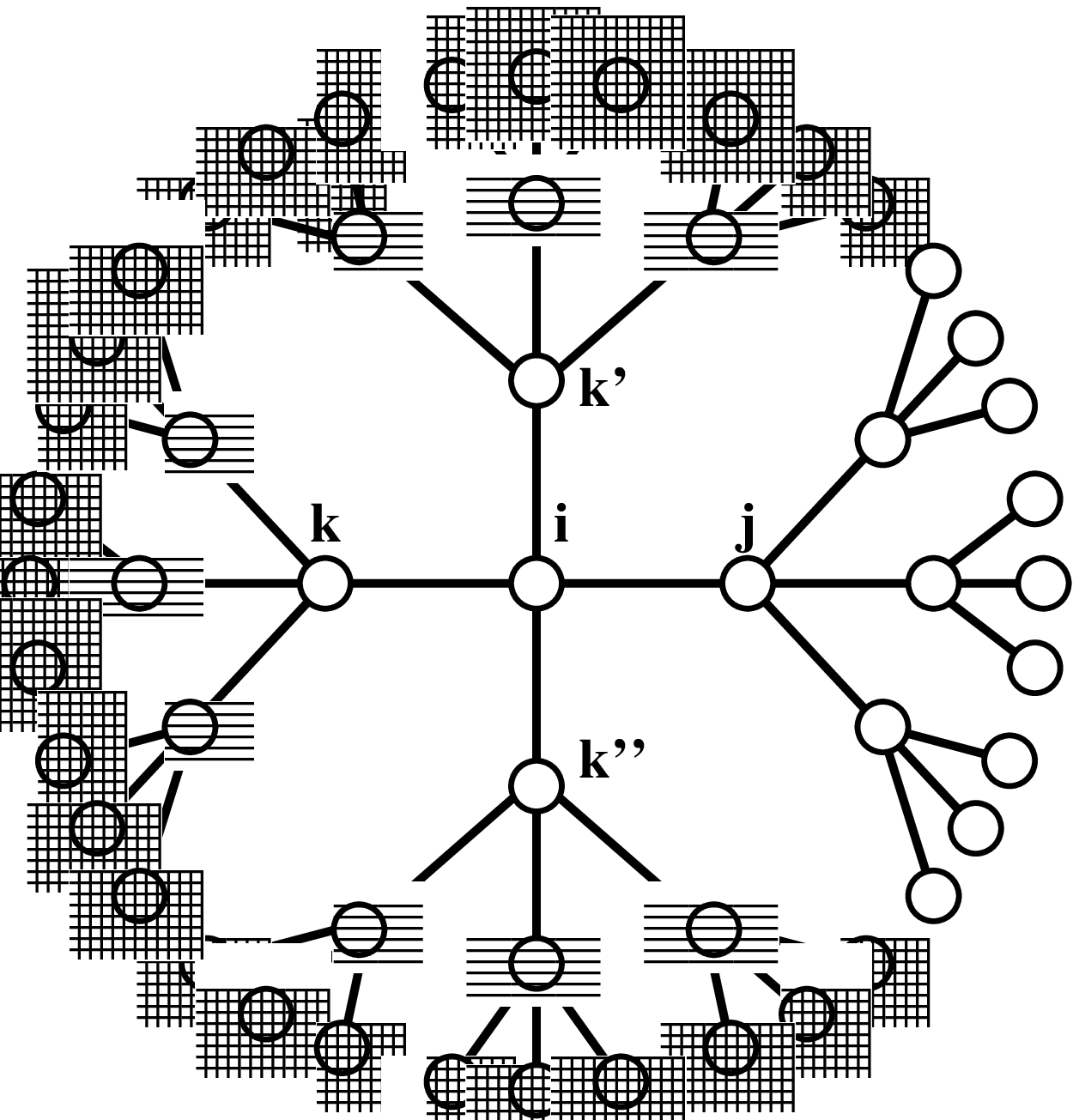}
    \caption{(a)~The NN sites of $i$, except $j$, are denoted by ${\cal Z}(i|j)$, 
      marked by horizontal shading. ${\cal B}(i|j)$ denotes the sites
      on the branches starting on any $k\in{\cal Z}(i|j)$ and leading away
      from $i$, marked by horizontal or double shading. $\Xi(i|j)$
      [Eq.~(\ref{eq:Xi-def})] involves a trace over shaded sites.
      (b)~Here the shaded sites are involved in the trace appearing in
      $\Xi(j|i)$ [Eq.~(\ref{eq:Xi-def})]. When combined with the sites
      traced over in $\Xi(i|j)$, i.e., the shaded sites in (a), one
      obtains a trace over all sites except $i$ and $j$, leading to
      the expression for the local Green function $G_{i}$ in
      Eq.~(\ref{eq:G-pathint2}).  (c) The shaded sites in (a) can also
      be enumerated by combining ${\cal B}(k|i)$ for all $k\in{\cal
        Z}(i|j)$, leading to Eq.~(\ref{eq:Xi-relation}) which
      expresses $\Xi(i|j)$ in terms of a product
      $\Xi(k|i)\Xi(k'|i)\cdots$.}
    \label{fig:branches}
  \end{figure}

  We now consider $\tonetwo$ hopping on the Bethe lattice as in
  Eq.~(\ref{eq:H-t1t2}).  In this case the following decomposition of
  the Bethe lattice turns out to be useful.  For two NN sites $i$ and
  $j$ we let ${\cal Z}(i|j)$ denote the set of NN sites of $i$, but
  with site $j$ omitted. Furthermore ${\cal B}(i|j)$ shall denote the
  sites on all the branches which begin at the sites in ${\cal
    Z}(i|j)$ and lead away from $i$. These definitions are illustrated
  in Fig.~\ref{fig:branches}a.

  Now consider a partial trace where we integrate out the Grassmann
  variables for the sites in ${\cal B}(i|j)$, i.e., all the sites
  connected to $i$ except for those on the branch starting at $j$. We
  thus define, for NN sites $i$ and $j$,
  \begin{align}
    \Xi(i|j)
    &=
    \int\prod_{k\in{\cal B}(i|j)}{\cal D}[\overline{c}_{k},c_{k}]
    \;\;\;
    e^{-S(i|j)}
    \,,
    \label{eq:Xi-def}
    \\
    S(i|j)
    &=
    \sum_{k\in{\cal B}(i|j)}
    \left[
    S_{\text{loc}}(k)
    +
    S_{\text{hop}}(k,i)
    +
    S_{\text{hop}}(k,j)
    \right]
    +
    \frac{1}{2}
    \sum_{k,m\in{\cal B}(i|j)}
    S_{\text{hop}}(k,m)
    \,,
  \end{align}
  i.e., we include only the part $S(i|j)$ of the action in the trace
  that involves Grassmann variables on the branches ${\cal B}(i|j)$.
  As a consequence, the only Grassmann variables remaining in
  $\Xi(i|j)$ are $\overline{c}_{i}$, $c_{i}$ and $\overline{c}_{j}$,
  $c_{j}$. Because they are connected to other $\overline{c}_{k}$,
  $c_{k}$ in $S(i|j)$ only linearly, $\Xi(i|j)$ is of Gaussian form,
  \begin{align}
    \Xi(i|j)
    &\propto
    \exp\left(
      -
      \Big(
      \overline{c}_i
      \;\;
      \overline{c}_j
      \Big)
      \bm{X}_{\gamma}
      \Big(
        \begin{array}{c}
          c_i\\
          c_j
        \end{array}
      \Big)
    \right)
    \,,
    &
    \bm{X}_{\gamma}
    &=\left(
      \begin{array}{cc}
        \Woo_{\gamma}  & \Woi_{\gamma}  \\
        \Wio_{\gamma}  & \Wii_{\gamma}
      \end{array}
    \right)
    \,,\label{eq:Xi-gaussian}
  \end{align}
  for $i\in\gamma$, $j\in\bamma$.  Here the coefficients in the
  exponent depend only on the sublattices to which $i$ and $j$
  belong, due to the translational invariance of the infinite Bethe
  lattice.  These coefficients will be determined below.

  The partial trace in Eq.~(\ref{eq:Xi-def}) serves two purposes. On
  the one hand the local Green function (\ref{eq:G-pathint1}) can now be
  written as a trace over $i$ and a neighboring site $j$, together
  with their respective partial traces and the remaining parts of the
  action,
  \begin{align}
    G_{i}(z)
    &=
    \frac{
      \int{\cal D}[\overline{c}_{i},c_{i},\overline{c}_{j},c_{j}]
      \;
      \Xi(i|j)
      \;
      \Xi(j|i)
      \;
      e^{-S_{\text{loc}}(i)-S_{\text{loc}}(j)-S_{\text{hop}}(i,j)}
      \;
      \overline{c}_{i}c_{i}
    }{
      \int{\cal D}[\overline{c}_{i},c_{i},\overline{c}_{j},c_{j}]
      \;
      \Xi(i|j)
      \;
      \Xi(j|i)
      \;
      e^{-S_{\text{loc}}(i)-S_{\text{loc}}(j)-S_{\text{hop}}(i,j)}
    }
    \,.\label{eq:G-pathint2}
  \end{align}
  This equals (\ref{eq:G-pathint1}) because, when considering the two
  neighboring sites $i$ and $j$, $\Xi(i|j)$ contains the contributions
  of all the sites ``behind'' $i$, whereas $\Xi(j|i)$ contains the
  contributions of all the sites ``behind'' $j$, as illustrated in
  Figs.~\ref{fig:branches}a and \ref{fig:branches}b, and the remaining
  action involves only $i$ and $j$ because the hopping is only between
  NN or NNN sites.  We note in passing that a generalization to
  hopping \emph{beyond NNN}, while possible in principle, will become
  increasingly complicated. Inserting Eq.~(\ref{eq:Xi-gaussian}) into
  (\ref{eq:G-pathint2}) gives a Gaussian integral which can be
  performed by completing the square, yielding
  \begin{align}
    G_{\gamma}(z)
    &=
    \frac{
      z_{\bamma}-\Woo_{\bamma}-\Wii_{\gamma}
    }{
      (z_{\gamma}-\Woo_{\gamma}-\Wii_{\bamma})
      (z_{\bamma}-\Woo_{\bamma}-\Wii_{\gamma})
      -
      (\tone+\Woi_{\bamma}+\Wio_{\gamma})
      (\tone+\Woi_{\gamma}+\Wio_{\bamma})
    }
    \label{eq:G-with-W}
  \end{align}
  for the local Green function on sublattice $\gamma$, with
  $z_{\gamma}=z-\epsilon_{\gamma}$.
  
  On the other hand one can obtain a functional equation for
  $\Xi(i|j)$ by decomposing the Bethe lattice in different ways.
  Recall that $\Xi(i|j)$ contains the branches which start at the NN
  sites $k$ of $i$, where $k$ $\neq$ $j$. We can also move one site
  away from $i$ and consider $\Xi(k|i)$, which contains the branches
  which start at $k$'s NN neighbors, except $i$, as illustrated in
  Fig.~\ref{fig:branches}c. We recover $\Xi(i|j)$ when we multiply the
  $\Xi(k|i)$'s, include pieces of the action which connect the ${\cal
    B}(k|i)$ among each other and with $i$ and $j$, and trace over the
  sites $k$. Thus we arrive at
  \begin{multline}
    \Xi(i|j)
    =
    \int\prod_{k\in{\cal Z}(i|j)}
    {\cal D}[\overline c_k,c_k]
    \;    
    \prod_{k\in{\cal Z}(i|j)}
    \Xi(k|i)      
    \\
    \;
    \times
    \;
    \exp\!\left(
      -
      \sum_{k\in{\cal Z}(i|j)}
      \left[
        S_{\text{loc}}(k)+S_{\text{hop}}(k,i)+S_{\text{hop}}(k,j)
        +
        \frac{1}{2}
        \sum_{k'(\neq k)\in{\cal Z}(i|j)}
        S_{\text{hop}}(k,k')
      \right]
    \right)\!
    \,,\label{eq:Xi-relation}
  \end{multline}
  where $i$ and $j$ are any two NN sites.
  
  Now $\Xi(i|j)$ can be determined from the above relation, using its
  Gaussian form (\ref{eq:Xi-gaussian}).  Due to the translational
  symmetry of the infinite Bethe lattice it suffices to consider, say,
  $i=0\in\gamma$ and $j=1\in\bamma$.  From
  Eqs.~(\ref{eq:Xi-gaussian}) and (\ref{eq:Xi-relation}) we then
  obtain
  \begin{align}
    \Xi(0|1)
    &\propto
    \int 
    \prod_{k=2}^{Z}
    {\cal D}[\overline{c}_k,c_k]
    \exp\!\left(
      -K\,\overline{c}_0 \Wii_{\bamma} c_0
      -\sum_{i=2}^{Z}(\overline{\xi}c_i+\overline{c}_i\xi)
      -\sum_{i,j=2}^{Z}\overline{c}_i M_{ij}c_j
    \right)\!
    \,,\label{eq:Xi-t1t2}
  \end{align}
  where $K=Z-1$,
  $\overline{\xi}=(\tone+\Wio_{\bamma})\overline{c}_0+\ttwo\overline{c}_1$,
  $\xi=(\tone+\Woi_{\bamma})c_0+\ttwo c_1$, and the product and sums
  are over nearest neighbors of site $0$ other than site $1$.  The
  $K\times K$ matrix $\bm{M}$ in the Gaussian integral and its inverse
  are of the form
  $M_{ij}=\delta_{ij}\,a+(1-\delta_{ij})\,b$ and
  $(\bm{M}^{-1})_{ij}=\delta_{ij}\,c+(1-\delta_{ij})\,d$,
  respectively, with $c=(a+(K-2)b)/q$, $d=-b/q$, and
  $q=a^2+(K-2)ab-(K-1)b^2$.  Here $a=\Woo_{\bamma}-z_{\bamma}$ and
  $b=\ttwo$. The Gaussian integration in Eq.~(\ref{eq:Xi-t1t2}) is
  performed by completing the square, yielding
  \begin{align}
    \Xi(0|1)
    &\propto
    \exp\!\left(
      -K\,\overline{c}_0 \Wii_{\bamma} c_0
      -\overline{\xi}\xi\sum_{i,j=2}^{Z}(\bm{M}^{-1})_{ij}
    \right)\!
    =
    \exp\left(
      -K\overline{c}_0\Wii_{\bamma}c_0
      -K\overline{\xi}\frac{a-b}{q}\xi
    \right)
    \,.\label{eq:Xi-result}
  \end{align}
  Comparing with Eq.~(\ref{eq:Xi-gaussian}) we thus obtain the
  following system of equations:
  \begin{align}
    \Woo_{\gamma}
    &=
    K\,\Wii_{\bamma}+K\,(\tone+\Wio_{\bamma})\,(\tone+\Woi_{\bamma})\,\widehat{G}_{\bamma}
    \,,
    \\
    \Woi_{\gamma}
    &=
    K\,\ttwo\,(\tone+\Wio_{\bamma})\,\widehat{G}_{\bamma}
    \,,
    \\
    \Wio_{\gamma}
    &=
    K\,\ttwo\,(\tone+\Woi_{\bamma})\,\widehat{G}_{\bamma}
    \,,
    \\
    \Wii_{\gamma}
    &=
    K\,\ttwosq\,\widehat{G}_{\bamma}
    \,,
  \end{align}
  with the abbreviation
  $\widehat{G}_{\gamma}=[z_{\gamma}-\Woo_{\gamma}-(K-1)\ttwo]^{-1}$.

  \subsection{Local Green function for arbitrary connectivity}
  
  For easier discussion of the limit $Z\to\infty$ we use the scaled
  hopping parameters of Eq.~(\ref{eq:scaling}). Putting
  $\tilde{z}=z+\txtwo/K$ and
  $\widetilde{G}_{\gamma}=\widehat{G}_{\gamma}/(1+\txtwo\widehat{G}_{\gamma})
  = (\tilde{z}-\epsilon_{\gamma}-\Woo_{\gamma})^{-1}$ we find, after
  some rearrangement,
  \begin{align}
    \Woo_{\gamma}
    &=
    \frac{
      \txonesq\,\widetilde{G}_{\bamma}\,(1-\txtwo\,\widetilde{G}_{\bamma})
    }{
      (1-\txtwo\,\widetilde{G}_{\gamma}-\txtwo\,\widetilde{G}_{\bamma})^2
    }
    +
    \frac{
      \txtwosq\,\widetilde{G}_{\gamma}
    }{
      1-\txtwo\,\widetilde{G}_{\gamma}
    }
    =
    \tilde{z}-\epsilon_{\gamma}-\widetilde{G}_{\gamma}^{-1}    
    \,,\label{eq:W00-solution}
  \end{align}
  \begin{align}
    \Woi_{\gamma}
    &=
    \Wio_{\gamma}
    =
    \frac{1}{\sqrt{K}}
    \frac{\txone\txtwo\,\widetilde{G}_{\bamma}}{1-\txtwo\,\widetilde{G}_{\gamma}-\txtwo\,\widetilde{G}_{\bamma}}
    \,,
    &
    \Wii_{\gamma}
    &=
    \frac{1}{K}
    \frac{\txtwosq\,\widetilde{G}_{\bamma}}{1-\txtwo\,\widetilde{G}_{\bamma}}
    \,.\label{eq:W11-solution}
  \end{align}
  For given $\tilde{z},\epsilon_{\text{A}},\epsilon_{\text{B}}$,
  Eq.~(\ref{eq:W00-solution}) is a system of two symmetric equations
  for $\widetilde{G}_{\text{A}},\widetilde{G}_{\text{B}}$ and we denote the
  appropriate solution by $\widetilde{G}_{\gamma} =
  f(\tilde{z},\epsilon_{\gamma},\epsilon_{\bamma})$; note that
  this function does not depend explicitly on $K$.  Inserting
  Eqs.~(\ref{eq:W00-solution})-(\ref{eq:W11-solution}) into
  (\ref{eq:G-with-W}) we can then express the local Green function as
  \begin{align}
    G_{\gamma}(z)^{-1}
    &=
    \widetilde{G}_{\gamma}^{-1}
    -
    \frac{1}{K}
    R(\widetilde{G}_{\gamma},\widetilde{G}_{\bamma})
    \,,
    \;\;\;\;\;\;\;\;\;\;\;\;\;\;\;\;\;\;\;\;\;\;\;\;\;\;\;\;\;\;
    \widetilde{G}_{\gamma}
    =
    f(z+\tfrac{\txtwo}{K},\epsilon_{\gamma},\epsilon_{\bamma})
    \,,\label{eq:G-solution}
    \\
    R(\widetilde{G}_{\gamma},\widetilde{G}_{\bamma})
    &=
    \frac{
      \txonesq\,\widetilde{G}_{\bamma}\,(1-\txtwo\,\widetilde{G}_{\bamma})
    }{
      (1-\txtwo\,\widetilde{G}_{\gamma}-\txtwo\,\widetilde{G}_{\bamma})^2(1-p\txtwo\,\widetilde{G}_{\bamma})
    }
    +
    \frac{\txtwo}{1-\txtwo\,\widetilde{G}_{\gamma}}
    \,,\label{eq:G-function}
  \end{align}
  where again $p=Z/K$.  Taking the limit $K\to\infty$ in
  Eq.~(\ref{eq:G-solution}) we find that $G\INF_{\gamma}(z) =
  f(z,\epsilon_{\gamma},\epsilon_{\bamma})$, and also
  $\lim_{K\to\infty}\widehat{G}_{\gamma}(z)=\GA(z)$
  [Eq.~(\ref{eq:rpe-Ghat-definition})].  
  This leads us to the remarkable conclusion that \emph{the local
    Green function $G_{\gamma}(z)$ for arbitrary $K$ is a rational
    function of the local Green functions $G\INF_{\gamma'}(z)$ for
    infinite $K$},
  \begin{align}
    G_{\gamma}(z)
    &=
    \left[
      G\INF_{\gamma}(z+\tfrac{\txtwo}{K})^{-1}
      -
      \frac{1}{K}
      R\big(
      G\INF_{\gamma}(z+\tfrac{\txtwo}{K}),
      G\INF_{\bamma}(z+\tfrac{\txtwo}{K})
      \big)
    \right]^{-1}
    \,.\label{eq:G-arbitraryK}
  \end{align}
  The function $G\INF_{\gamma}$ is determined by
  Eq.~(\ref{eq:W00-solution}). However, this is clearly the same
  implicit equation that was obtained in
  (\ref{eq:rpe-F-expression}) and (\ref{eq:rpe-F-result}), thus confirming
  the RPE calculation of the previous section.  The function
  $G\INF_{\gamma}(z)$ is obtained in the next section.
  
  Finally we note that the unexpected relation (\ref{eq:G-arbitraryK})
  can be checked for only NN hopping,
  \begin{align}
    g_{\gamma}(z)
    &=
    \left[
      g\INF_{\gamma}(z)^{-1}
      -
      \frac{\txonesq}{K}\,
      g\INF_{\bamma}(z)
    \right]^{-1}
    \,,
    &&&
    (\ttwo&=0)
  \end{align}
  which is indeed fulfilled by the results (\ref{eq:nn-only}) for
  this case.

  \section{Topological approach}\label{sec:operatormethod}

  \subsection{Operator identities for hopping Hamiltonians}
  
  Recently a topological approach to the tight-binding spectrum of
  $H_{\text{kin}}$ on the Bethe lattice was developed
  \cite{eckstein04}, which we will extend to the case of additional
  A-B on-site energies $\Hloc$ here.
  
  We begin with a short review of the method of
  Ref.~\cite{eckstein04}.  Unlike crystal lattices, the (infinite)
  Bethe lattice has the property that the number of paths between two
  lattice sites $i$ and $j$ which consist of $n$ NN steps depends only
  on the topological distance $\distance{i}{j}$, but not on the
  relative orientation of $i$ and $j$. This ``distance regularity''
  entails polynomial relations \cite{fiol} among the tight-binding
  Hamiltonians $H_d$.  For the Bethe lattice they are given by
  \cite{eckstein04}
  \begin{align}%
    U_n(\tilde{H_1}/2)
    &=
    \sum_{s=0}^{\floor{n/2}}
    \frac{\tilde{H}_{n-2s}}{K^s}
    \,,
    &&&&&
    \tilde{H}_d
    &=
    U_d(\tilde{H_1}/2)
    -\frac{1}{K}\,U_{d-2}(\tilde{H}_1/2)
    \,,
    &&
    (d\geq2)
    \label{eq:chebychev}
  \end{align}%
  where $U_n(x)$ are the Chebychev polynomials of the second kind
  \cite{abramowitz84a}.
  Similar relations involving the Hermite polynomials hold for the
  infinite-dimensional  hypercubic lattice \cite{blumer,blumer2}.
  By contrast, Eq.~(\ref{eq:chebychev}) is valid for any $K$,
  including the one-dimensional chain ($K$ $=$ $1$) as well as the
  limit $K\to\infty$.  These relations can also be expressed by means
  of the generating function
  \begin{align}
    \frac{1-x^2}{1-xH_1+Kx^2}
    &=
    \sum_{d=0}^{\infty}H_d\,x^d
    \,,
    &
    \frac{1-x^2/K}{1-x\tilde{H}_1+x^2}
    &=
    \sum_{d=0}^{\infty}\tilde{H}_d\,x^d
    \,.\label{eq:geometricseries}
  \end{align}
  From these operator identities one concludes that for the Bethe
  lattice the eigenstates of any hopping Hamiltonian
  (\ref{eq:H-general}) are the same as those of $\tilde{H}_1$, and its
  eigenvalues $\epsilon(\lambda)$ can be expressed as a function of
  the eigenvalues $\lambda$ of $\tilde{H}_1$. The calculation of
  spectral properties, such as the density of states \cite{eckstein04}
  from this effective dispersion $\epsilon(\lambda)$ is then
  straightforward. The method works well for arbitrary hopping
  $t_{d}$ since no explicit enumerations are required.
  
  We now incorporate the effect of alternating on-site
  energies [Eq.~(\ref{eq:H-loc})],
  \begin{align}
    \Hloc
    &=
    \sum_{\gamma=\text{A},\text{B}}
    \epsilon_\gamma
    \sum_{i\in\gamma}
    \ket{i}\bra{i}
    =
    \bar{\epsilon}
    +
    \epsilon
    \,V
    \,,
    &
    V
    &=
    \sum_i
    (-1)^i\ket{i}\bra{i}
    \,,
    &
    \frac{\epsilon_{\text{A}}\pm\epsilon_{\text{B}}}{2}
    &=:
    \left\{
      \begin{array}{l}
        \bar{\epsilon}\\
        \epsilon
      \end{array}
    \right.
    \,,
  \end{align}
  where $(-1)^i=\pm1$ for $i\in\text{A},\text{B}$, e.g.,
  $(-1)^i=(-1)^{\distance{i}{j}}$ for some fixed site $j\in\text{A}$.
  Since the operators $H_{2d}$ ($H_{2d+1}$) connect the same
  (different) sublattices they commute (anticommute) with $V$,
  respectively,
  \begin{align}
    VH_d-(-1)^dH_dV
    &=
    0
    \,,
    &
    VH_1^n-(-1)^nH_1^nV
    &=
    0
    \,,\label{eq:commutators}
  \end{align}
  where the second equation follows from the polynomial relations
  (\ref{eq:chebychev}) between $H_d$ and $H_1^n$. Together with
  $V^2=1$ we immediately obtain the useful new operator identity
  \begin{align}
    (\alpha V+\beta H_1)^{2n}
    &=
    (\alpha^2 +\beta^2 H_1^2)^{n}
    \,,
    &
    n
    &=
    0,1,2,\ldots
    \,.\label{eq:V-identity}
  \end{align}
  for arbitrary constants $\alpha$, $\beta$. This identity makes it
  possible to reduce resolvent operators involving $V$ and $H_1$ to
  simpler expressions.
  
  As a simple application let us consider only NN hopping
  (\ref{eq:nn-only}).  We find by straightforward series expansion and
  partial fraction decomposition
  \begin{align}
    \frac{1}{z-\HNN}
    &=
    \frac{
      z-\bar{\epsilon}+\epsilon V+\txone\tilde{H}_1
    }{
      (z-\bar{\epsilon})^2-(\epsilon^2+\txonesq\tilde{H}_1^2)
    }
    =
    \frac{1}{2}
    \sum_{s=\pm1}
    \left(1+\frac{z-\bar{\epsilon}+\epsilon V}{s\sqrt{x}}\right)
    \frac{1}{s\sqrt{x}-\txone \tilde{H}_1}
    \,,
  \end{align}
  again with $x=(z-\epsilon_{\text{A}})(z-\epsilon_{\text{B}})$.  We
  therefore conclude that for NN hopping, surprisingly, any Green
  function for $\epsilon_{\text{A}}\neq\epsilon_{\text{B}}$ can be
  written as a linear combination of two Green functions for the
  homogeneous case, $\epsilon_{\text{A}}=\epsilon_{\text{B}}$, at
  other arguments.
  
  \subsection{Results for $\subsectionmath{\tonetwo}$ hopping}
  
  For $\tonetwo$ hopping [Eq.~(\ref{eq:H-t1t2})] there are now two
  possible routes to the local Green function $G_{\gamma}(z)$ for
  arbitrary $K$. On one hand, according to
  Eq.~(\ref{eq:G-arbitraryK}) of the previous section, $G_{\gamma}(z)$
  is determined by $G\INF_{\gamma}(z)$ alone.  The latter was already
  obtained in Ref.~\cite{eckstein04} from the operator identity
  (\ref{eq:geometricseries}) for the homogeneous case, together with a
  diagonalization of the 2x2 sublattice problem.  This gave
  \begin{subequations}%
    \label{eq:top-G-inf}%
    \begin{align}%
    G\INF_{\gamma}(z)
    &=
    \frac{1}{2\txtwo}
    +
    \frac{
      1
      }{
        2(\lambda_2^2-\lambda_1^2)\txtwosq
      }
    \sum_{i=1}^2
    \frac{
      (-1)^i
      [
      z_{\bar{\gamma}}
      -
      (\lambda_i^2-1)\txtwo
      ]
      \sqrt{\lambda_i-2}\sqrt{\lambda_i+2}
    }{
      \lambda_i
    }
    \,,
    \label{eq:green-AB-ttprime}
  \end{align}%
  \begin{align}%
    \lambda_{1,2}
    &=
    \sqrt{\calA\pm\sqrt{\calA^2-\calB}}
    \,,
    &
    \calA
    &=
    1+\frac{(z_{A}-z_{B})\txtwo+\txonesq}{2\txtwosq}
    \,,
    &
    \calB
    &=
    \bigg[\frac{z_{A}}{\txtwo}+1\bigg]\bigg[\frac{z_{B}}{\txtwo}+1\bigg]
    \,,
  \label{eq:lambda-AB}
  \end{align}%
  \end{subequations}%
  where $z_{\gamma}=z-\epsilon_{\gamma}$ and all square roots are
  given by their principal branches. While $G_{\gamma}(z)$ is given
  explicitly by Eqs.~(\ref{eq:G-arbitraryK}) and (\ref{eq:top-G-inf}),
  we note that this approach would be less promising for hopping
  beyond NNN.
  
  On the other hand we may directly use the operator identities
  (\ref{eq:chebychev}) and (\ref{eq:V-identity}), which provide the
  relation $\tilde{H}_2=\tilde{H}_1^2-p= (vV+\tilde{H}_1)^2-v^2-p$,
  with $v$ arbitrary and $p=Z/K$.  This gives us
  \begin{align}
    H
    &=
    \Hloc
    +
    \txone\tilde{H}_1
    +
    \txtwo\tilde{H}_2
    =
    \bar{\epsilon}
    -
    \txtwo\,(p+v^2)
    +
    \txone\,(vV+\tilde{H}_1)
    +
    \txtwo\,(vV+\tilde{H}_1)^2
    \,,
  \end{align}
  with $v=\epsilon/\txone$. Performing the partial fraction
  decomposition for the resolvent we arrive at
  \begin{subequations}%
    \label{eq:top-G-allK}%
    \begin{align}%
      \frac{1}{z-H} &= \frac{\xi_1+\xi_2}{\xi_1-\xi_2}
      \sum_{k=1}^{2}\frac{(-1)^k}{\xi_k+\bar{\epsilon}-\HNN} \,, &
      G_{\gamma}(z) &= \frac{\xi_1+\xi_2}{\xi_1-\xi_2}
      \sum_{k=1}^{2}(-1)^kg_{\gamma}(\xi_k+\bar{\epsilon}) \,,
    \end{align}
    \begin{align}
      \xi_{1,2} &= \frac{ -\txonesq \pm
        \sqrt{\txonequ+4\txone^2\txtwo\,(z-\bar{\epsilon})+4\txtwosq\,(\epsilon^2+p\txonesq)}
      }{2\txtwo} \,,
    \end{align}%
  \end{subequations}%
  where $g_{\gamma}(z)$ again denotes the local Green function
  (\ref{eq:nn-only}) for only NN hopping.  We note that this
  remarkably short route to $G_{\gamma}(z)$ is straightforward to
  carry out also for hopping \emph{beyond NNN}, as well as for
  off-diagonal Green functions.

  \section{Results for the local Green function}\label{sec:green}
  
  Using the results of the previous sections, i.e.,
  Eqs.~(\ref{eq:G-arbitraryK}) and (\ref{eq:top-G-inf}), or
  Eqs.~(\ref{eq:top-G-allK}) and (\ref{eq:nn-only}), the local Green
  function is now available for arbitrary $\tonetwo$ hopping and
  on-site energies $\epsilon_{\text{A,B}}$ on the Bethe lattice for
  finite or infinite coordination number.
  
\begin{figure}[tbp]
    \centering
    \includegraphics[clip,width=0.495\textwidth,angle=0]{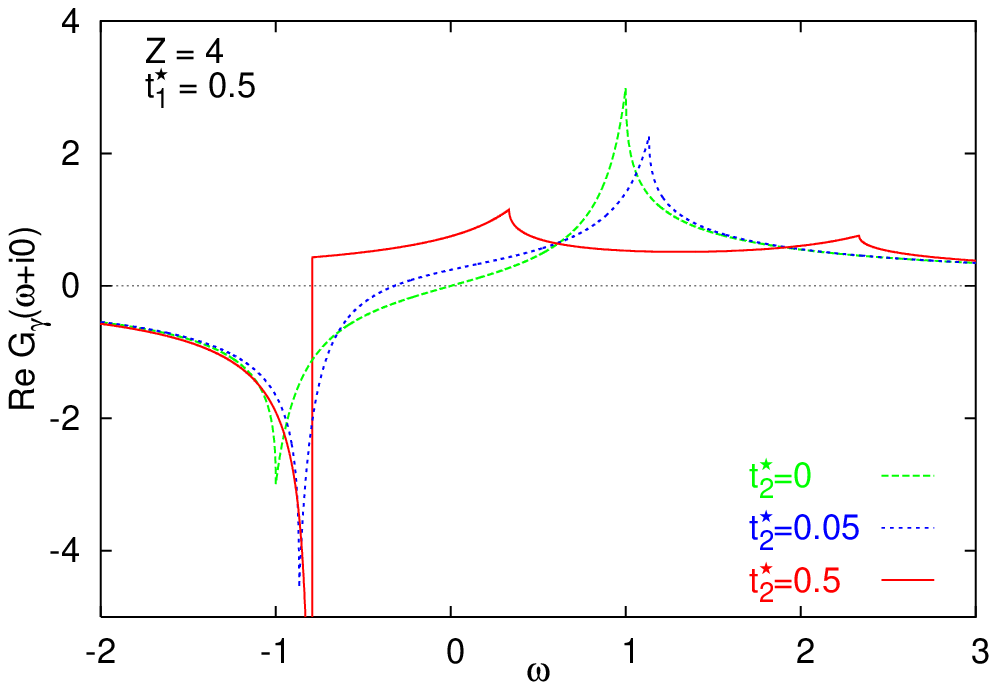}%
    \hfill%
    \includegraphics[clip,width=0.495\textwidth,angle=0]{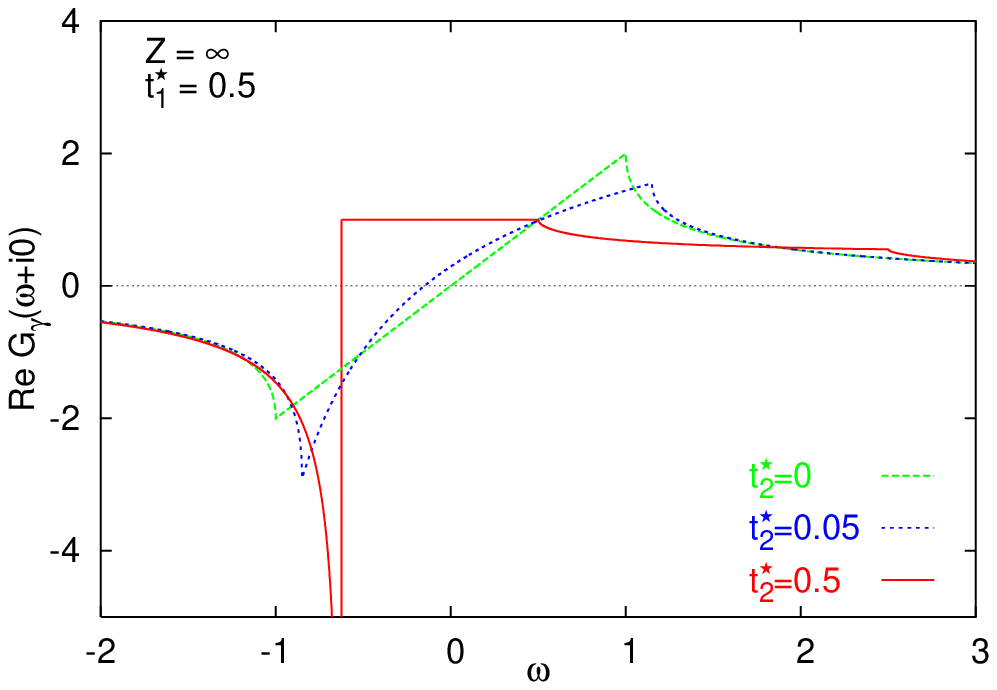}
    \\
    \includegraphics[clip,width=0.495\textwidth,angle=0]{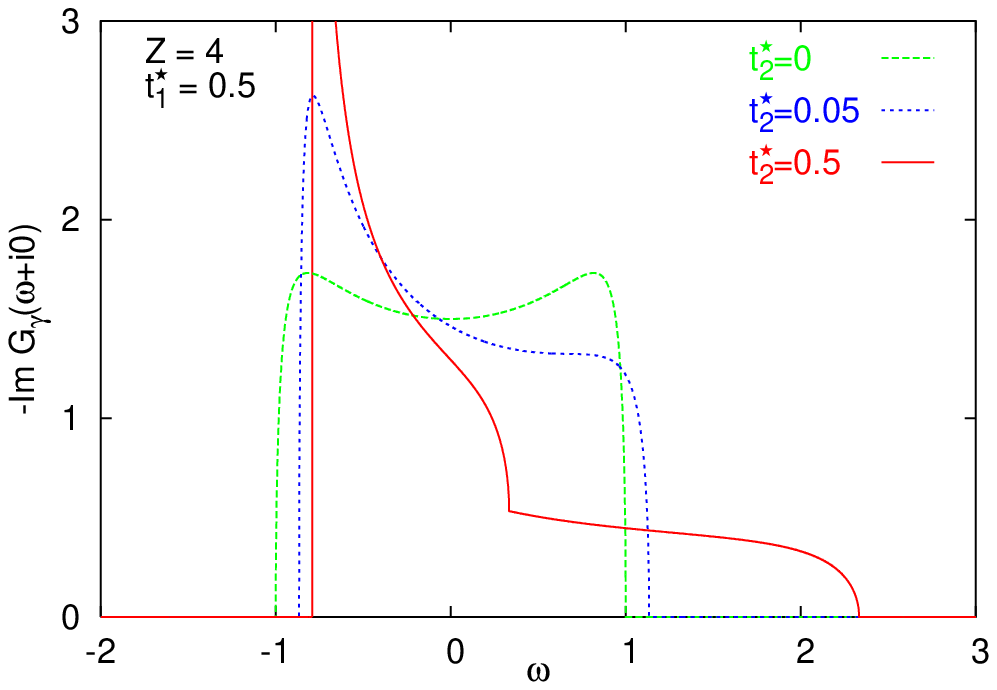}%
    \hfill%
    \includegraphics[clip,width=0.495\textwidth,angle=0]{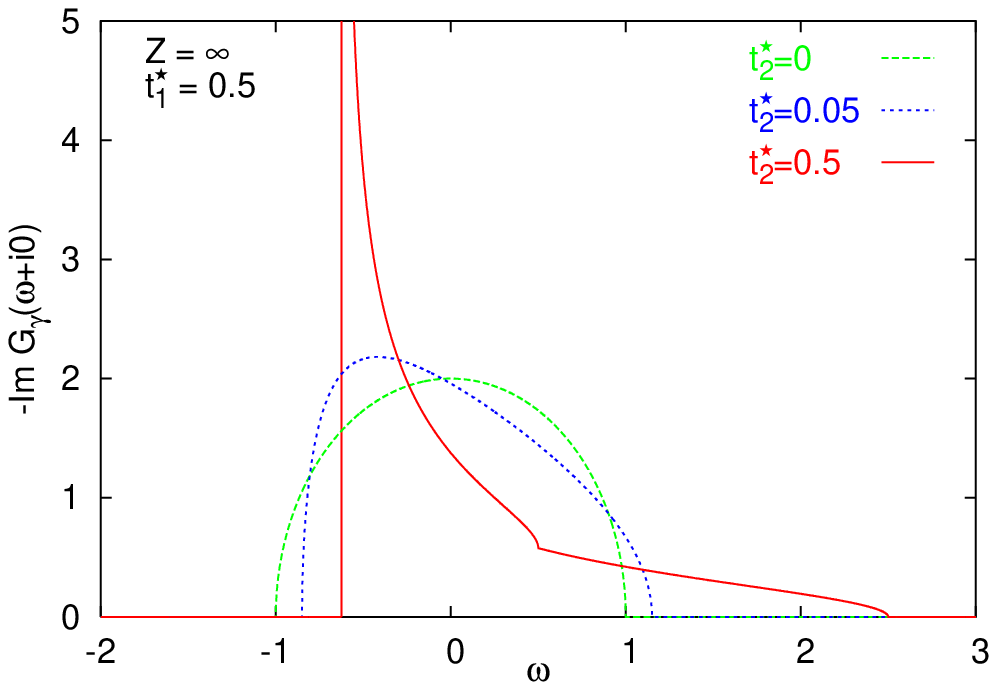}
    \caption{Local Green function $G_{\gamma}(\omega+i0)$
      for the Bethe lattice with
      $\epsilon_{\text{A}}=\epsilon_{\text{B}}=0$, $\txone=0.5$, and
      $\txtwo/\txone=0,\frac{1}{10},1$. Left column: $Z=4$. Right
      column: $Z=\infty$. Vertical lines mark divergences.}
    \label{fig:green-hom}
  \end{figure}

\begin{figure}[tbp]
    \centering
    \includegraphics[clip,width=0.495\textwidth,angle=0]{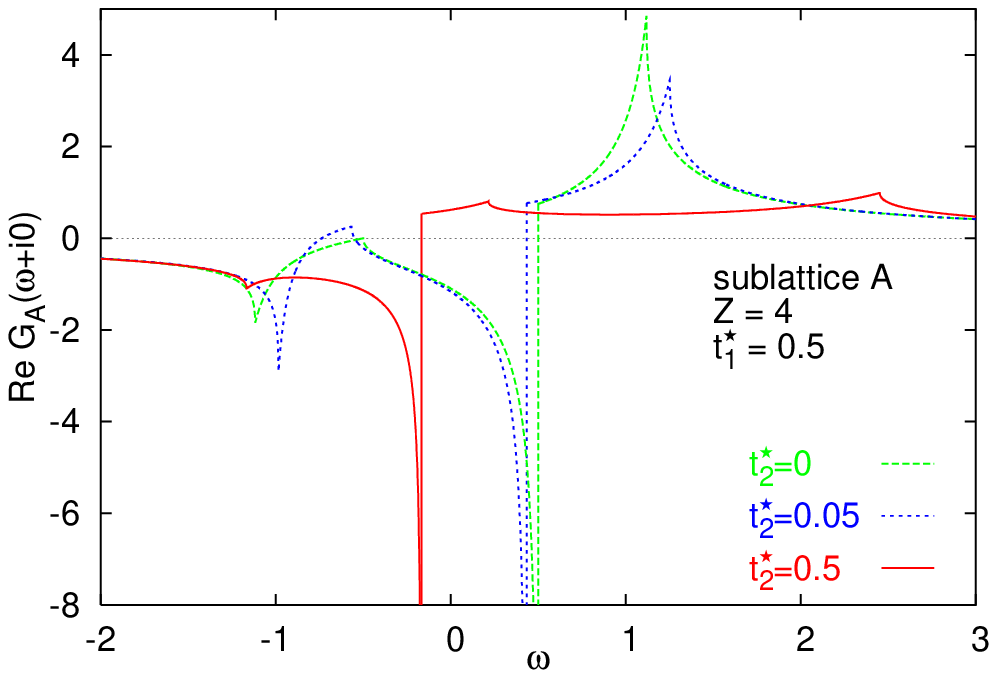}%
    \hfill%
    \includegraphics[clip,width=0.495\textwidth,angle=0]{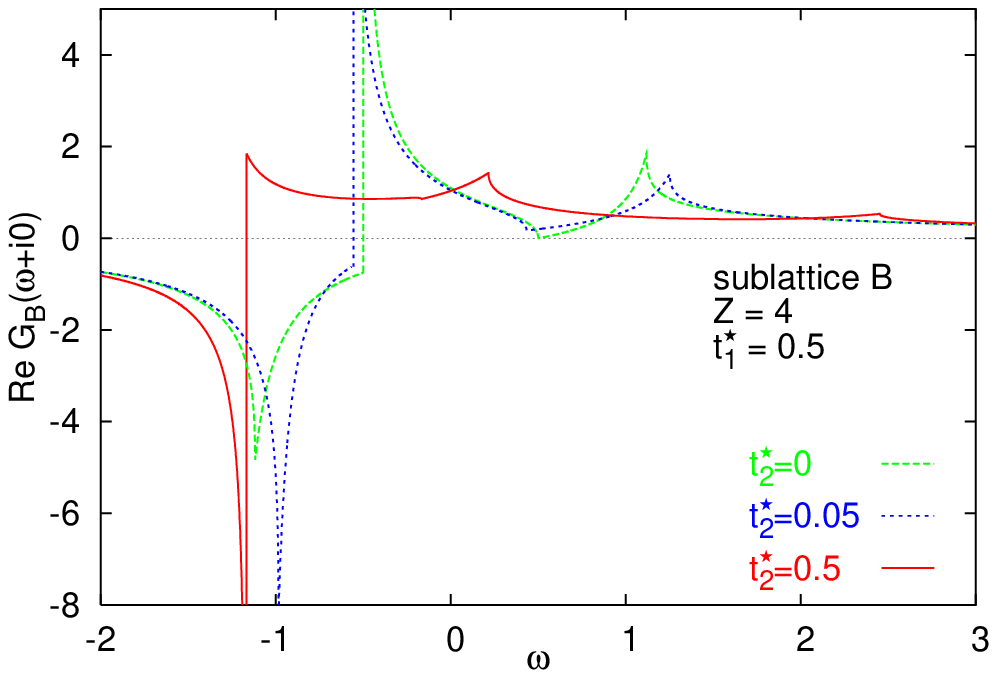}%
    \\
    \includegraphics[clip,width=0.495\textwidth,angle=0]{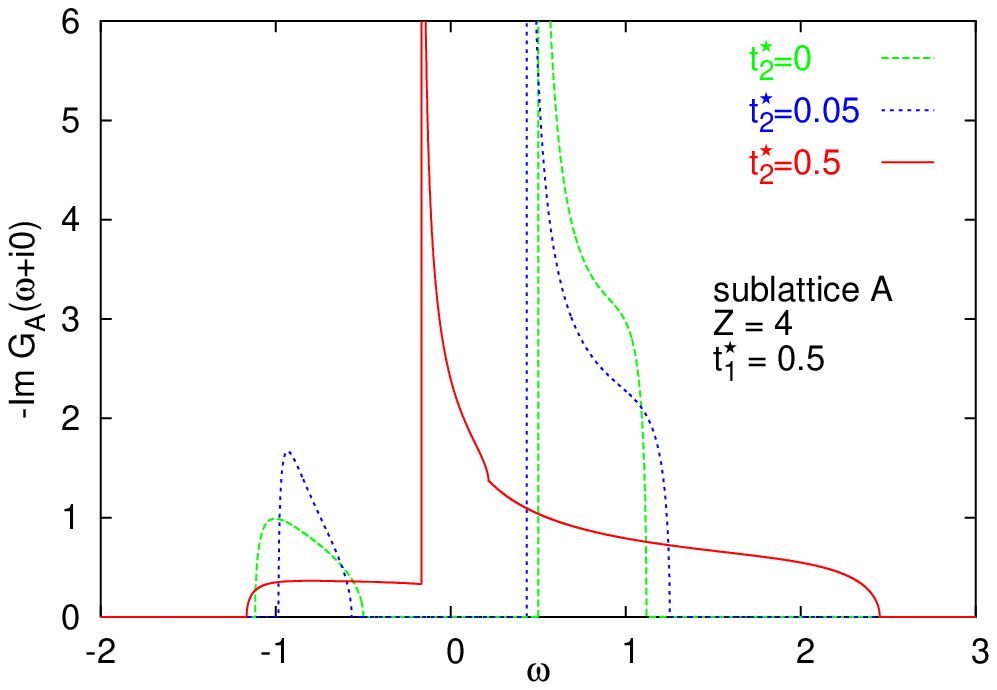}%
    \hfill%
    \includegraphics[clip,width=0.495\textwidth,angle=0]{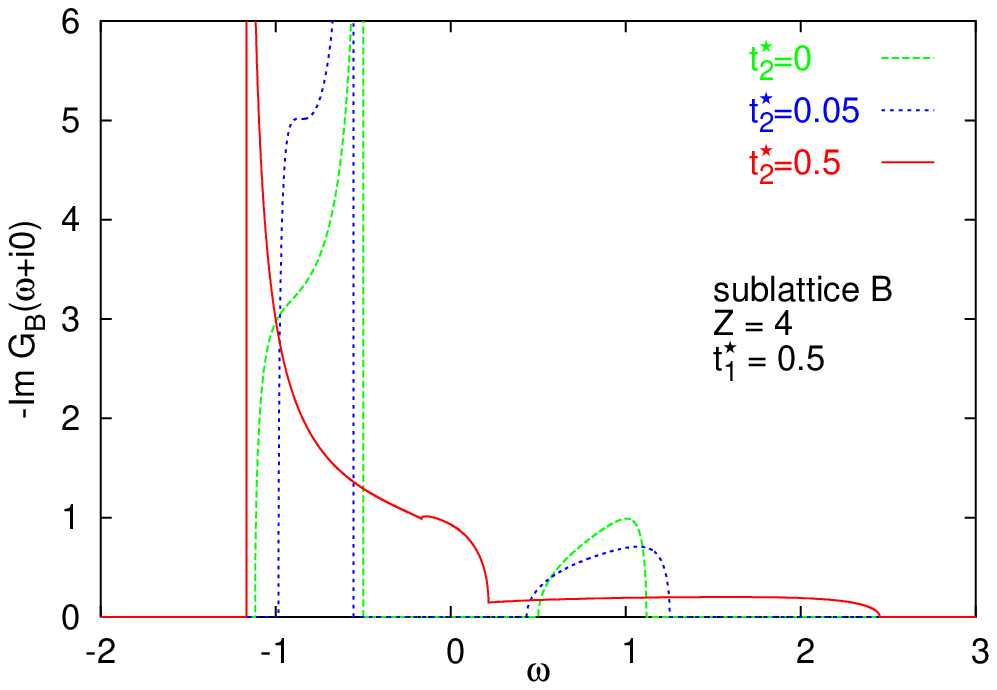}%
    \caption{Local Green function $G_{\gamma}(\omega+i0)$ 
      for the Bethe lattice with $Z=4$,
      $\epsilon_{\text{A}}=-\epsilon_{\text{B}}=\txone=0.5$ and
      $\txtwo/\txone=0,\frac{1}{10},1$.  Left column: Local Green
      function for sublattice A, $G_{\text{A}}(\omega+i0)$, which for
      $\txtwo=0$ also appears in Ref.~\cite{economou}.  Right column:
      Local Green function for sublattice B,
      $G_{\text{B}}(\omega+i0)$; note the small cusp at
      $\omega=-0.167$.  Vertical lines mark divergences.}
    \label{fig:green-alt-fin}
  \end{figure}

\begin{figure}[tbp]
    \centering
    \includegraphics[clip,width=0.495\textwidth,angle=0]{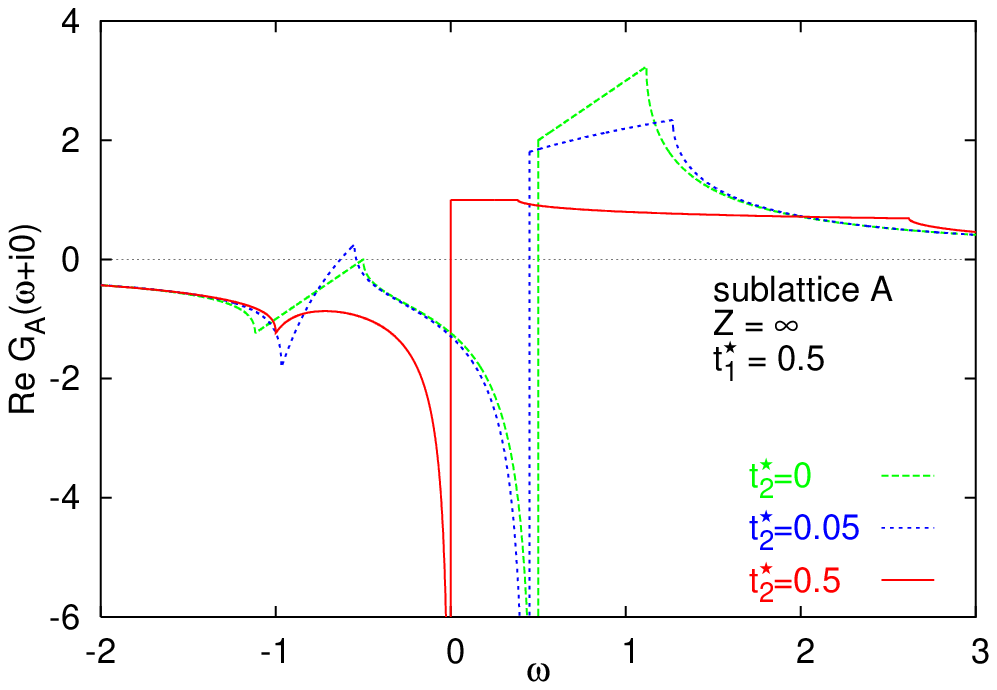}%
    \hfill%
    \includegraphics[clip,width=0.495\textwidth,angle=0]{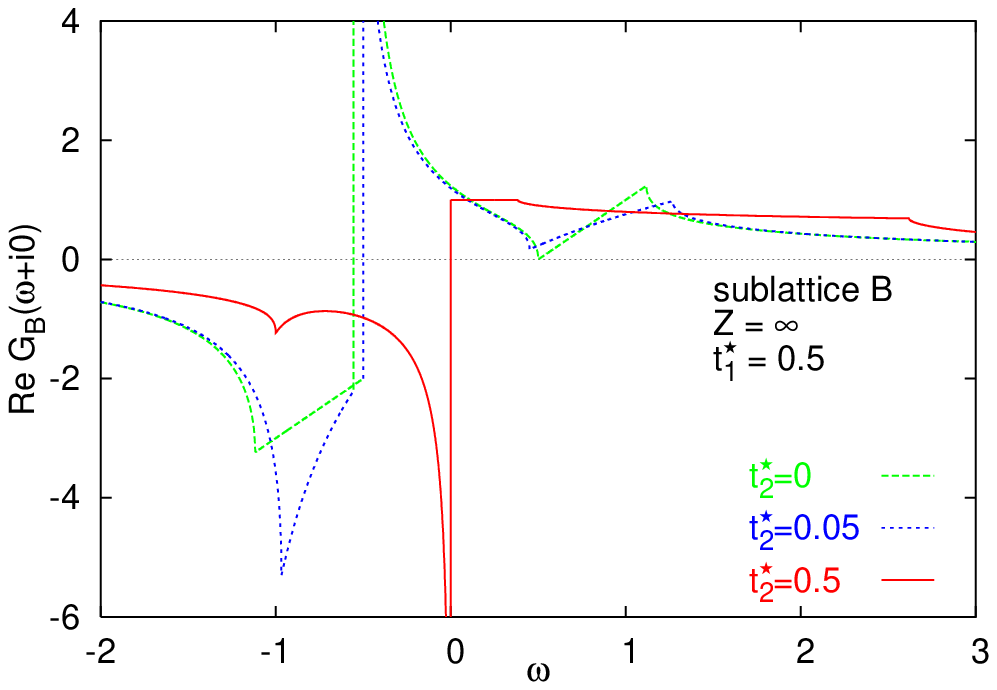}%
    \\
    \includegraphics[clip,width=0.495\textwidth,angle=0]{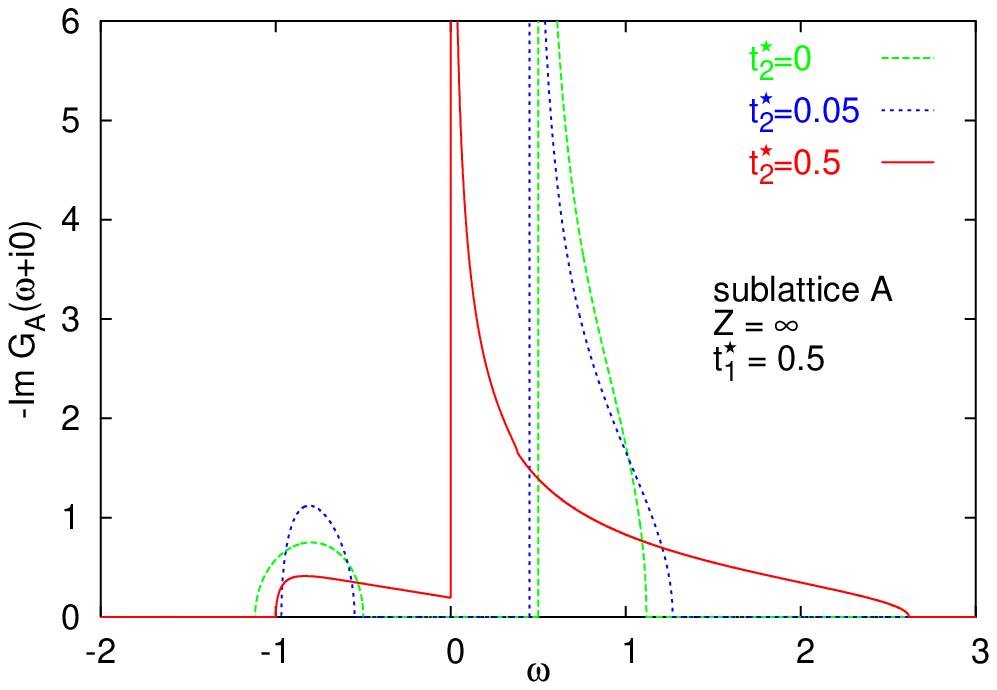}%
    \hfill%
    \includegraphics[clip,width=0.495\textwidth,angle=0]{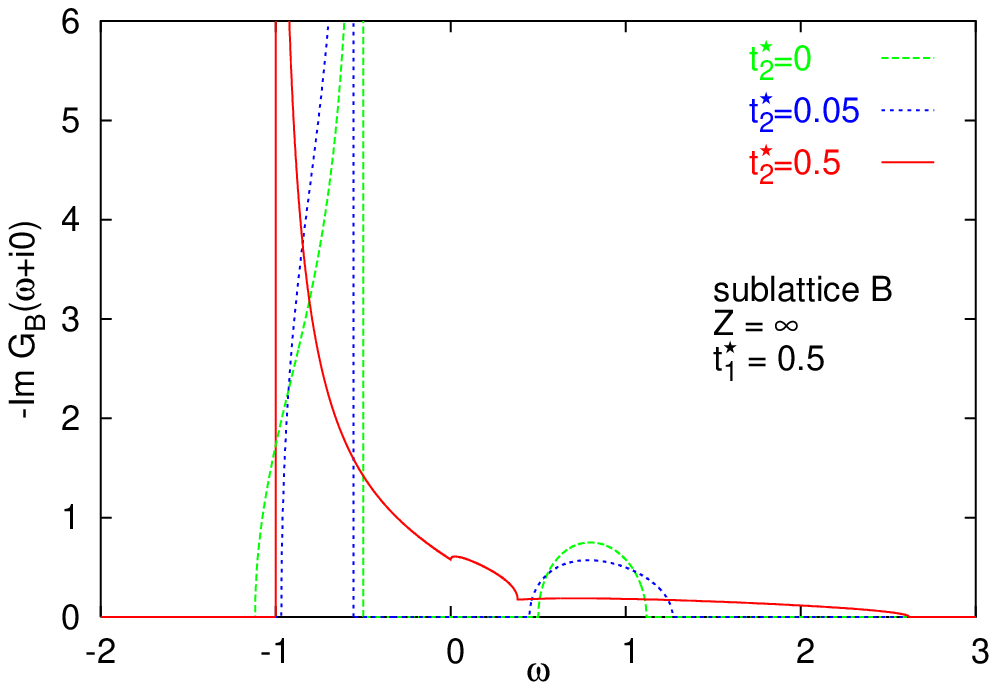}%
    \caption{Same as Fig.~\ref{fig:green-alt-fin}, but for $Z=\infty$.}
    \label{fig:green-alt-inf}
  \end{figure}

  In Figs.~\ref{fig:green-hom}-\ref{fig:green-alt-inf} we consider
  both $Z=4$ and $Z=\infty$ and compare the unfrustrated case
  ($\txtwo=0$) to weak frustration ($\txtwo=\txone/10$) and strong
  frustration ($\txtwo=\txone$).  The homogeneous case,
  $\epsilon_{\text{A}}=\epsilon_{\text{B}}$, is shown in
  Fig.~\ref{fig:green-hom}, whereas we chose
  $\epsilon_{\text{A}}-\epsilon_{\text{B}}=2\txone$ in
  Figs.~\ref{fig:green-alt-fin} and \ref{fig:green-alt-inf}.

  From these spectra several effects of frustration may be observed.
  Beginning with the homogeneous case (Fig.~\ref{fig:green-hom}) the
  imaginary part of the Green function, i.e., the density of states,
  is no longer symmetric if $\txtwo\neq0$, as noted in
  Sec.~\ref{sec:rpe}. This is the expected generic behavior for a
  bipartite lattice with hopping between the same sublattices.
  Furthermore, as discussed already in Ref.~\cite{eckstein04}, a
  square-root singularity at one band edge develops for large enough
  $|\txtwo|$.  For strong frustration one notices the appearance of
  several additional cusps in both real and imaginary part of the
  Green function, as well as an increase in bandwidth.  It is also
  apparent that in the limit $Z=\infty$ the Green function loses some
  of its features. In this case its real part is linear or even flat
  in part of the band.
  
  These characteristics persist for the case
  $\epsilon_{\text{A}}\neq\epsilon_{\text{B}}$
  (Figs.~\ref{fig:green-alt-fin} and \ref{fig:green-alt-inf}).  In
  addition the symmetry $G_{\gamma}(z)=G_{\bamma}(-z)$ is absent for
  $\txtwo\neq0$.  For strong frustration more van-Hove
  singularities appear. We also note that the band gap, which is
  present for only NN hopping due to the alternating on-site energies,
  is closed for large enough NNN hopping.
  
  In summary, the Green function for $\tonetwo$ hopping shows several
  qualitatively new properties, which are likely to have an impact
  also on its behavior in interacting many-body or disordered systems,
  in particular for site-diagonal disorder.

  \section{Conclusion}\label{sec:conclusion}
  
  Due to the special topology of the Bethe lattice, the calculation of
  the Green function of a quantum-mechanical particle for hopping
  beyond nearest neighbors seemed untractable so far.  In this paper
  we presented the derivation of an explicit expression for the local
  Green function for $\tonetwo$ hopping and sublattice-dependent
  on-site energies $\epsilon_{\text{A}},\epsilon_{\text{B}}$ for
  arbitrary coordination number $Z$, employing a set of complementary
  analytical techniques.  Implicit equations for $G\INF_{\gamma}$ were
  derived by RPE. They also follow from a path integral approach,
  which furthermore yielded the local Green function $G_{\gamma}$ for
  arbitrary $Z$ as a rational function of $G\INF_{\gamma}$.  It should
  be noted that such a functional relation between Green functions at
  different $Z$ is quite unexpected and to our knowledge does not
  occur for any other lattice.  From a topological approach explicit
  expressions for $G_{\gamma}$ and $G\INF_{\gamma}$ were obtained in
  terms of the Green function $g_{\gamma}$ for only NN hopping.  We
  found that NNN hopping makes the density of states asymmetric and
  may induce additional van-Hove singularities, increase the total
  bandwidth, and close gaps that were opened by alternating on-site
  energies. From the experience of these results we conclude that it
  will be worthwhile to investigate the effects which hopping beyond
  nearest neighbors has on the physics of interacting many-body and
  disordered systems on the Bethe lattice.

  \begin{acknowledgement}
    This work was supported in part by Sonderforschungsbereich 484 of
    the Deutsche Forschungsgemeinschaft (M.K., M.E., K.B., D.V.).
    K.B.\ is supported in part by grant KBN~2~P03B~08224.  V.D.\ and
    D.T.\ are supported by the NSF through grant NSF-0234215.  G.K.\ 
    is supported by DOE DE-FG02-99ER45761.
  \end{acknowledgement}

\end{document}